\documentclass[a4paper,11pt]{article}
\pdfoutput=1 
\usepackage{jheppub} 
\usepackage[T1]{fontenc} 
\usepackage{color}

\def \be {\begin{equation}}
\def \ee {\end{equation}}
\def \ba {\begin{aligned}}
\def \ea {\end{aligned}}
\def \bea {\begin{eqnarray}}
\def \eea {\end{eqnarray}}

\title{Generalized free energy and dynamical state transition of the dyonic AdS black hole in the grand canonical ensemble}

\author[a,b]{Conghua Liu,}
\author[c]{Ran Li,}
\author[b]{Kun Zhang}
\author[d,*]{and Jin Wang \note[*]{Corresponding author.}}
\affiliation[a]{College of Physics, Jilin University, Changchun 130022, China}
\affiliation[b]{State Key Laboratory of Electroanalytical Chemistry, Changchun Institute of Applied Chemistry, Chinese Academy of Sciences, Changchun 130022, China}
\affiliation[c]{Department of Physics, Qufu Normal University, Qufu 273165, China}
\affiliation[d]{Department of Chemistry and Department of Physics and Astronomy, State University of New York at Stony Brook, Stony Brook, New York 11794, USA}
\emailAdd{jin.wang.1@stonybrook.edu}

\abstract{We study the generalized free energy of the dyonic AdS black hole in an ensemble with varying electric charge $q_E$ and fixed magnetic charge $q_M$. When we adjust the temperature $T$ and the electric potential $\Phi_E$ of the ensemble, the Ricci scalar curvature $R$ and electromagnetic potential $A_u$ usually diverge at the horizon. We regularize them and incorporate the off-shell corrections into the Einstein-Hilbert action. Alternatively, we find that the off-shell corrections can also be obtained by adding a boundary near the horizon to exclude the singularities. Ultimately, we derive the generalized free energy which is consistent with the definition of the thermodynamic relations. Based on the generalized free energy landscape, we can describe the dynamics of state transition as a stochastic process quantified by the Langevin equation. The path integral framework can be formulated to derive the time-dependent trajectory of the order parameter and the time evolution of the transition probability. By comparing the probability with the result of the classical master equation, we attribute the contribution to the probability of one pseudomolecule or antipseudomolecule (the instanton and anti-instanton pairs) to the rate of state transition. These results are consistent with the qualitative analysis of the free energy landscape.}

\begin{document}

\maketitle

\flushbottom

\section{Introduction}

Since the establishment of the four black hole thermodynamic laws~\cite{AA, AB, AC, AD}, phase transition of the AdS black hole has attracted much attention in the past decades. Two famous examples are the Hawking-Page phase transition which can be interpreted as the confinement/deconfinement phase transition in the context of AdS/CFT correspondence~\cite{AE, AF}, and the charged AdS black hole phase transition which was found to be similar to the liquid/gas phase transition~\cite{AG, AH}. Recently, the study of phase transition has been generalized to the extended phase space, where the cosmological constant is interpreted as the thermodynamic pressure to hold the consistency between the first law of black hole thermodynamics and the Smarr relation from the scaling argument~\cite{AI, AJ, AK, AL, AM, AN, AO,AP, AQ, ZAA, ZAB}. In the extended phase space, the similarities between the charged AdS black hole phase transition and the liquid/gas phase transition become more complete with similar equations of state and the same critical exponents~\cite{AL}. Furthermore, many novel phenomena such as reentrant phase transition~\cite{AM}, triple point~\cite{AN, AO}, multiple critical points~\cite{AP}, and $\lambda$-line phase transition~\cite{AQ} have been observed in the extended phase space.

Although there are extensive studies on the phase behaviors in different black hole systems, the dynamics of state transition have not been investigated adequately until very recently. A recent study suggests that the thermal fluctuations play a vital role in the state transition, and the dynamics can be described as a stochastic process quantified by the Fokker-Planck equation~\cite{AR,AS}. Such a viewpoint has attracted significant attention and has been extended to studies on various black hole systems~\cite{AT,AU,AV,AW,AX,AY,AZ,BA,BB,BC,BD,BE}. We should note that we have used the term “state transition” in our paper instead of “phase transition” in previous studies~\cite{AR,AS,AT,AU,AV,AW,AX,AY,AZ,BA,BB,BC,BD,BE}, because it better describes the dynamical process. In the case of a first-order phase transition, it occurs at a specific temperature when two locally stable states have the same free energy. However, the state transition is not limited solely to the phase transition temperature. The globally stable state, which has lower free energy, still has the probability to switch to the locally stable state with higher free energy due to thermal fluctuations. In other words, state transitions have a chance to occur at other temperatures as well.

As we know, the Langevin equation is an equivalent description of the Fokker-Planck equation, and the dynamics of state transition can also be quantified by the Langevin equation. In~\cite{BF}, we utilized the Langevin equation to formulate the path integral framework and investigated the dynamics of charged AdS black hole state transition in the canonical ensemble. Compared with the method of solving the Fokker-Planck equation, there are three advantages. Firstly, we can quantify the path showing visually how the state transition proceeds. The unstable transition states can be easily identified on the path as they do not have resident time. Secondly, our framework can give the analytical expression of the time evolution of the probability rather than the numerical results. Thirdly, the transition rate between the two stable states has a clear significance, it is the contribution to the probability of one pseudomolecule or antipseudomolecule. 

Whether we use the description of the Fokker-Planck equation or the Langevin equation, the driving force of the stochastic process is provided by the generalized free energy landscape. In the generalized free energy landscape, the free energy is a continuous function of the order parameters (See Fig.~\ref{Fig1} for the generalized free energy landscape). Only the extreme points on the landscape correspond to the on-shell black hole states whose manifolds are regular, all other states are off-shell states whose manifolds are conical singular. Therefore, the standard definition of free energy should be generalized to apply not only to the on-shell states but also to the off-shell states. In addition to composing the free energy landscape, the off-shell states also serve as intermediate transition states that reveal the process of the path during the state transition. Thus, it is necessary for us to introduce the off-shell states and the generalized free energy. Recently, generalized free energy has another interesting application in black hole thermodynamics, where the authors have used generalized free energy to treat black hole solutions as topological defects~\cite{BG}. One can see Ref.~\cite{BH, BI, BJ, BK, BL, BM, BN, BO, BP, BQ, BR, BS, BT, BU, BV, BW, BX, BY} for the latest studies.

Initially, the generalized free energy is defined by the thermodynamic relations for the canonical ensemble~\cite{AR, AS}. Considering the AdS black hole as a system in contact with a thermal bath located at infinity, the temperature of the canonical ensemble is treated as an external parameter that can be adjusted arbitrarily. However, the black hole is on-shell only when the ensemble temperature is equal to the Hawking temperature. The generalized free energy can then be obtained by replacing the Hawking temperature $T_H$ in the standard definition of the free energy $G=M-T_HS$ with the ensemble temperature $T$. Subsequently, a more concrete and solid foundation for the generalized free energy in the canonical ensemble has been derived by utilizing the Hawking-Gibbons gravitational path integral on the Euclidean manifold with a conical singularity~\cite{BZ}.

However, previous studies on the generalized free energy and the dynamical state transition have mainly focused on the canonical ensemble, with the black hole radius being the only order parameter. In this paper, we will investigate the generalized free energy and the dynamical state transition of the dyonic AdS black hole in the grand canonical ensemble using the Langevin equation. We will keep the magnetic charge fixed while varying the electric charge of the dyonic AdS black hole, making the electric charge and the radius the chosen order parameters. Although a very recent study shows that the generalized free energy in a grand canonical ensemble can be obtained by the Legendre transformation of the generalized free energy in the canonical ensemble~\cite{CA}, we aim to find a more fundamental derivation of the generalized free energy in the grand canonical ensemble, i.e., starting from the Euclidean action and using the Hawking-Gibbons gravitational path integral. We provide two methods for calculating the Euclidean action and its related generalized free energy. These approaches involve the regularization of divergences at the horizon and the addition of a boundary to exclude the singularities, respectively. Interestingly, both approaches produce identical outcomes. Furthermore, the framework of the dynamical state transition in the grand canonical ensemble differs from that in the canonical ensemble, and our studies are presented as follows.


\section{The on-shell thermodynamics of the dyonic AdS black hole}
The action of the Einstein-Maxwell theory in four-dimensional AdS space can be written as
\be
\ba
\label{eq:1}
I_{bulk}=-\frac{1}{16\pi}\int_M d^4x\sqrt{-g}(R-2\Lambda-F^2)-\frac{1}{8\pi}\int_{\Sigma_+}d^3x\sqrt{|h|}K,
\ea
\ee
where the Hawking-Gibbons boundary term is added for a well-defined action principle~\cite{CB}. $h$ is the determinant of the induced metric on the boundary $\Sigma_+$ at infinity, and $K=h^{\mu\nu}K_{\mu\nu}$ is the trace of the extrinsic curvature of $\Sigma_+$ as embedded in $M$.

The variation of the action in Eq.~(\ref{eq:1}) can be calculated as
\be
\ba
\label{eq:2}
\delta I_{bulk}=&-\frac{1}{16\pi}\int_Md^4x\sqrt{-g}[R_{\mu\nu}-\frac{1}{2}g_{\mu\nu}(R-2\Lambda)-(2F_{\mu}^{\rho}F_{\nu}{\rho}-\frac{1}{2}g_{\mu\nu}F_{\alpha\beta}F^{\alpha\beta})]\delta g^{\mu\nu}\\
&-\frac{1}{4\pi}\int_Md^4x\sqrt{-g}(\nabla_{\mu}F^{\mu\nu})\delta A_{\nu}+\frac{1}{4\pi}\int_{\Sigma_+}d^3x\sqrt{|h|}n_{\mu}F^{\mu\nu}\delta A_{\nu},
\ea
\ee
where $n_{\mu}$ in the last term is the outward pointing unit normal vector of the boundary $\Sigma_+$ at infinity. As we can see, the first two terms provide the equations of motion for the gravitational field and the electromagnetic field. To have a well-defined action principle, we must impose $\delta A_{\mu}=0$ on the boundary $\Sigma_+$ to eliminate the last term. Therefore, Eq.~(\ref{eq:1}) can be used to study the ensemble with a fixed electric potential and a fixed magnetic charge.

With static spherical symmetry, the electromagnetic gauge potential and the metric of the dyonic AdS black hole can be obtained by solving the equations of motion for the electromagnetic and gravitational fields. This will yield:
\be
\ba
\label{eq:999}
A_{\mu}=-(\frac{q_{E}}{r}-\Phi_{EH})dt+q_M(1-\cos{\theta})d\phi
\ea
\ee
and
\be
\ba
\label{eq:3}
ds^2=-f(r)dt^2+\frac{1}{f(r)}dr^2+r^2d\theta^2+r^2\sin^2\theta d\phi^2,
\ea
\ee
with
\be
\ba
\label{eq:4}
f(r)=1+\frac{r^2}{l^2}-\frac{2M}{r}+\frac{q_E^2+q_M^2}{r^2}.
\ea
\ee
Here, $q_E$ and $q_M$ are the electric and magnetic charges, $M$ is the mass, $l^2=-\frac{3}{\Lambda}$ and $\Phi_{EH}=\frac{q_E}{r_h}$ is the on-shell electric potential. In Eq.~(\ref{eq:999}), the first and second terms have been chosen with the gauges which are regular on the horizon and the axis $\theta=0$, respectively.

The horizon radius $r_h$ is determined by $f(r_h)=0$, so the mass $M$ can be expressed by $r_h$ as
\be
\ba
\label{eq:5}
M=\frac{r_h}{2}+\frac{r_h^3}{2l^2}+\frac{q_E^2+q_M^2}{2r_h}.
\ea
\ee

The Hawking temperature is given by 
\be
\ba
\label{eq:6}
T_H=\frac{1}{4\pi}f'(r)|_{r=r_h}=\frac{1}{4\pi r_h}(1+\frac{3r_h^2}{l^2}-\frac{q_E^2+q_M^2}{r_h^2}).
\ea
\ee

In the extended phase space, the cosmological constant is considered as a variable that is related to pressure $P$ by $P=-\frac{\Lambda}{8\pi}$~\cite{AI, AJ, AK, AL}. Furthermore, to associate the charged AdS black hole with the van der Waals fluids, the specific volume $v$ of the charged AdS black hole is identified as $v=2r_h$~\cite{AL}. This identification is also valid for the dyonic AdS black hole, Eq.~(\ref{eq:6}) can then be rewritten as the equation of state
\be
\ba
\label{eq:7}
P=\frac{T_H}{v}-\frac{1-\Phi_{EH}^2}{2\pi v^2}+\frac{2q_M^2}{\pi v^4},
\ea
\ee
where we have used $\Phi_{EH}=\frac{q_E}{r_h}$. Eq.~(\ref{eq:7}) is similar to the equation of state for van der Waals fluids, as well as to the equation of state for charged AdS black hole. The critical point $(P_c,v_c, T_c)$ of the dyonic AdS black hole can be found by solving $\frac{\partial P}{\partial v}=0$ and $\frac{\partial^2 P}{\partial^2 v}=0$, which yields
\be
\ba
\label{eq:8}
P_c=\frac{(1-\Phi_{EH}^2)^2}{96\pi q_M^2},\quad v_c=\frac{2\sqrt{6}q_M}{\sqrt{1-\Phi_{EH}^2}},\quad T_c=\frac{(1-\Phi_{EH}^2)^{\frac{3}{2}}}{3\sqrt{6}\pi q_M}.
\ea
\ee
Below the critical point, a phase transition of the liquid-gas type can occur.

We then proceed to study the action in Eq.~(\ref{eq:1}). In the framework of Hawking-Gibbons gravitational path integral, an analytic continuation $t\to i\tau$ can be used to derive the Euclidean action $I_E$, which is related to the partition function $Z_{grav}$ through the saddle point approximation as~\cite{CB}
\be
\ba
\label{eq:9}
I_E=-\ln Z_{grav}.
\ea
\ee
In other words, the Euclidean action can be used to derive the thermodynamic quantities in the ensemble. 

In Appendix~\ref{Cts} and~\ref{Bs}, we have shown two methods to calculate the Euclidean action. The first method is called ``counterterm subtraction", where the counterterm is added to cancel the divergence at infinity~\cite{CC, CD, CE, CF}. The second method is called ``background subtraction", which involves calculating the difference in Euclidean action between the dyonic AdS black hole and the pure AdS space~\cite{CG, CH, CI}. The pure AdS space is identified as turning off the mass, electric charge, and magnetic charge of the dyonic AdS black hole. Furthermore, the metric of AdS space is adjusted so that it matches that of the dyonic AdS black hole at a cutoff distance $\Tilde{R}$, and we finally take the limit of $\Tilde{R}\to\infty$. In this method, the Hawking-Gibbons boundary term is canceled out and the counterterm is not necessary.

We should note that what we calculate in the Appendix is $I_{M/\Sigma_{-}}$ at an arbitrary temperature $T$ and electric potential $\Phi_E$ rather than at the Hawking temperature $T_H$ and the on-shell potential $\Phi_{EH}$. $I_{M/\Sigma_-}$ is the action of the Euclidean manifold $M$ excluding the surface $\Sigma_-$ located at the horizon, i.e, $I_{M/\Sigma_-}=I_{M}-I_{\Sigma_-}$. Because $T$ and $\Phi_E$ are arbitrary, $I_M$ is now called the reduced action~\cite{CJ,CK}. In the next section, we will see that the off-shell corrections are $I_{\Sigma_-}$, such a term vanishes for the on-shell black hole solutions. Thus, when $T=T_H$ and $\Phi_E=\Phi_{EH}$, the reduced action $I_M$ will recover the on-shell action $I_{os}$, which is also equal to $I_{M/\Sigma_{-}}$. As we can see in Appendix~\ref{Cts} and~\ref{Bs}, both methods yield the same result as
\be
\ba
\label{eq:10}
I_{M/\Sigma_-}=\frac{\beta r_h}{4}-\frac{\beta r_h^3}{4l^2}-\frac{\beta q_E^2}{4r_h}+\frac{3\beta q_M^2}{4r_h},
\ea
\ee
where $\beta=\frac{1}{T}$ is the period of imaginary time $\tau$.

Taking $\beta=\beta_H$ and $\Phi_E=\Phi_{EH}$ in Eq.~(\ref{eq:10}), we can obtain $I_{os}$. By substituting $I_{os}$ into Eq.~(\ref{eq:9}), we can derive the partition function $Z_{grav}$. Then, the free energy for the on-shell black hole can be calculated as 
\be
\ba
\label{eq:11}
G_{os}&=\frac{I_{os}}{\beta_H}\\
&=\frac{r_h}{4}-\frac{r_h^3}{4l^2}-\frac{q_E^2}{4r_h}+\frac{3q_M^2}{4r_h}\\
&=(\frac{r_h}{2}+\frac{r_h^3}{2l^2}+\frac{q_E^2+q_M^2}{2r_h})-\frac{1}{4\pi r_h}(1+\frac{3r_h^2}{l^2}-\frac{q_E^2+q_M^2}{r_h^2})\pi r_h^2-\frac{q_E^2}{r_h}\\
&=M-T_HS-\Phi_{EH}q_E,
\ea
\ee
where $S=\pi r_h^2$ is the entropy of the black hole. It should be noted that the free energy $G_{os}$ is only effective for the on-shell black hole solutions. 

\begin{figure}[t]
\centering
\includegraphics[width=0.68\textwidth]{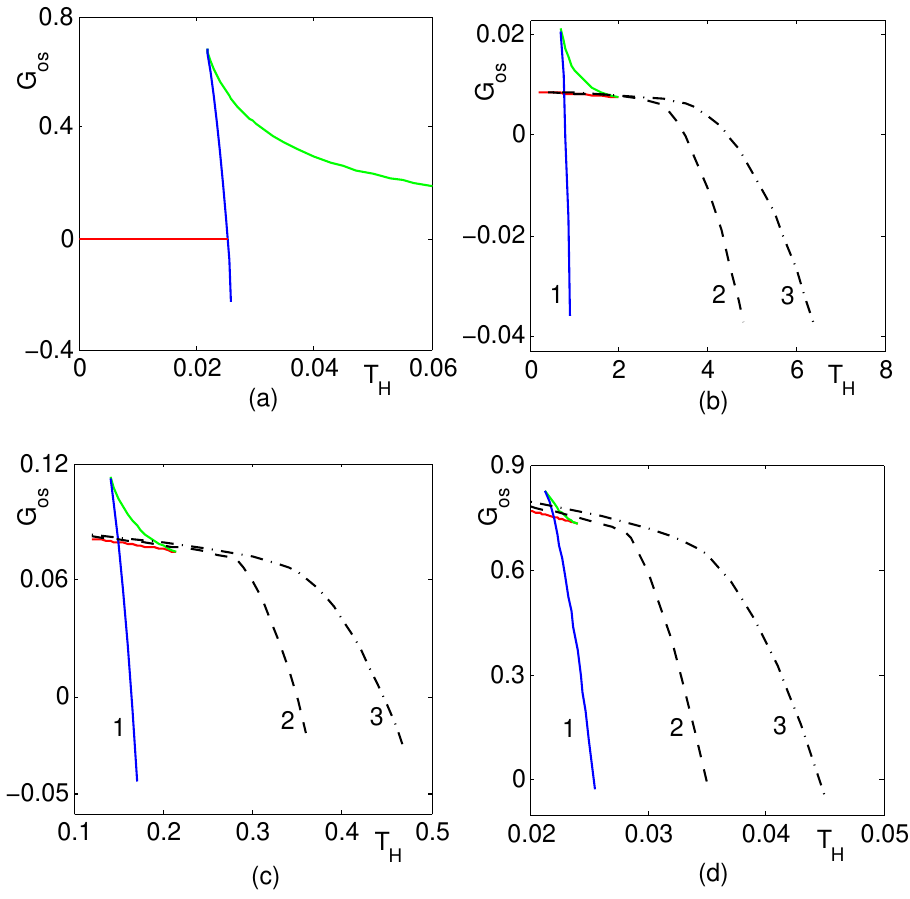}
\caption{The figures show the relationship between the free energy $G_{os}$ and the Hawking temperature $T_H$ for $\Phi_{EH}=0.5$, with varying magnetic charge in each subfigure labeled as: (a)$q_M=0$, (b)$q_M=0.01$, (c)$q_M=0.1$, and (d)$q_M=1$. In subfigure (a), the red, green, and blue lines represent the thermal AdS space, the unstable small black hole, and the stable large black hole, respectively. In subfigure (b), (c), and (d), the red, green, and blue lines represent the stable small black hole, the unstable intermediate black hole, and the stable large black hole, respectively. The pressures $P$ for curves $1$, $2$, and $3$ differ as follows: $P<P_c$ for curve $1$, $P=P_c$ for curve $2$, and $P>P_c$ for curve $3$.}
\label{Fig100}
\end{figure}

In Fig.~\ref{Fig100}, we have plotted the free energy $G_{os}$ as a function of the Hawking temperature $T_H$ for different magnetic charges $q_M$. When $q_M>0$ and $P$ is below the critical pressure $P_c$, three black hole branches can appear: a stable small black hole, an unstable intermediate black hole, and a stable large black hole. There can also be a first-order phase transition between the small and large black holes, similar to that of van der Waals fluids. As we decrease the magnetic charge, the free energy of the stable small black hole generally decreases to zero. When $q_M=0$, i.e. the dyonic AdS black hole recovers to the charged AdS black hole, the stable small black hole will disappear and become the thermal AdS space with a free energy equal to $0$. Then, there will only be two black hole branches, which can also be seen from Eq.~(\ref{eq:7}) by setting $q_M=0$. Thus, the phase transition between the small and large black holes becomes the Hawking-Page phase transition between the thermal AdS space and the large black hole. Actually, such a Hawking-Page phase transition can occur at any positive pressure. In our paper, we will focus on the case where $q_M>0$. Without loss of generality, we set $q_M=0.1$ in the following paper. In section~\ref{CaDi}, we will give some discussions about the case of $q_M=0$.

As mentioned in the introduction, in order to describe the dynamics of the black hole state transition, we should generalize the definition of the on-shell free energy so that it is also applicable to the off-shell black holes. The generalization is directly from the perspective of thermodynamics. We can consider the dyonic AdS black hole as a system in contact with a thermal and particle bath located at infinity in AdS space. Therefore, the temperature and the potential of the ensemble are external parameters that can be adjusted arbitrarily. By replacing the on-shell quantities $T_H$ and $\Phi_{EH}$ in Eq.~(\ref{eq:11}) with the ensemble parameters $T$ and $\Phi_E$, we can obtain the generalized free energy as
\be
\ba
\label{eq:12}
G_{gen}&=M-TS-\Phi_{E}q_E\\
&=\frac{r_h}{2}+\frac{r_h^3}{2l^2}+\frac{q_E^2+q_M^2}{2r_h}-T\pi r_h^2-\Phi_E q_E.
\ea
\ee
Here, $T$ and $\Phi_E$ are now free parameters and represent two physical degrees of freedom. If we take the derivative of $G_{gen}$ with respect to $r_h$ and $q_E$, both equal to $0$, we can recover the on-shell black hole solutions and lose the two degrees of freedom $T$ and $\Phi_{E}$. In the following section, we will derive the generalized free energy landscape from a more fundamental perspective by adding the off-shell corrections of $T\neq T_H$ and $\Phi_E\neq\Phi_{EH}$ to the Euclidean action.

\section{The off-shell corrections}
 The Euclidean metric of the dyonic AdS black hole is written as\footnote{We will use the analytic continuation $t\to i\tau$ in the whole paper.}
\be
\ba
\label{eq:13}
ds^2=f(r)d\tau^2+\frac{1}{f(r)}dr^2+r^2d\theta^2+r^2\sin^2{\theta}d\phi^2,
\ea
\ee
where
\be
\ba
\label{eq:14}
f(r)=1+\frac{r^2}{l^2}-\frac{2M}{r}+\frac{q_E^2+q_M^2}{r^2}.
\ea
\ee
As we know, the period of the imaginary time $\tau$ is always equal to the inverse of the ensemble temperature. If the ensemble temperature $T$ is not equal to the Hawking temperature $T_H$, a conical singularity will appear at the horizon~\cite{CL,CM,CN,CO}. To better understand this, we introduce a transformation of coordinates $d\rho=\frac{1}{\sqrt{f(r)}}dr$, which can be rewritten as
\be
\ba
\label{eq:15}
\rho=\int\frac{1}{\sqrt{f(r)}}dr.
\ea
\ee
Then, we calculate the Taylor expansion of $f(r)$ near the horizon $r_h$ as follows:
\be
\ba
\label{eq:16}
f(r)|_{r\to r_h}&=f(r_h)+f'(r_h)(r-r_h)+...\\
&\approx f'(r_h)(r-r_h),
\ea
\ee
where we have used $f(r_h)=0$ and take the approximation up to first order. 

Substituting Eq.~(\ref{eq:16}) into Eq.~(\ref{eq:15}) and imposing $\rho=0$ at $r=r_h$, we can obtain
\be
\ba
\label{eq:17}
\rho=\frac{2(r-r_h)^{1/2}}{\sqrt{f'(r_h)}}.
\ea
\ee

Substituting Eq.~(\ref{eq:17}) into Eq.~(\ref{eq:16}), we can obtain
\be
\ba
\label{eq:18}
f(r)|_{r\to r_h}=\frac{f'(r_h)^2\rho^2}{4}=\frac{(2\pi\rho)^2}{\beta_H^2},
\ea
\ee
where $\beta_H=\frac{4\pi}{f'(r_h)}$ is actually the inverse of the Hawking temperature $T_H$ in Eq.~(\ref{eq:6}).

Thus, the Euclidean metric of the dyonic AdS black hole near the horizon can now be rewritten as
\be
\ba
\label{eq:19}
ds^2&=\frac{(2\pi\rho)^2}{\beta_H^2}d\tau^2+d\rho^2+r^2(\rho)d\theta^2+r^2(\rho)\sin^2\theta d\phi^2\\
&=\rho^2d\xi^2+d\rho^2+r^2(\rho)d\theta^2+r^2(\rho)\sin^2\theta d\phi^2,
\ea
\ee
where $\xi=\frac{2\pi\tau}{\beta_H}$. 

As mentioned in the previous section, the temperature $T$ and the electric potential $\Phi_E$ of the ensemble are the external adjustable parameters. If we arbitrarily adjust the ensemble temperature, it corresponds to a change in the period $\beta$ of imaginary time $\tau$ given by $\beta=\frac{1}{T}$. When $\beta=\beta_H$, the $(\rho,\xi)$ space has the topology of a disk and the manifold is regular. When $\beta\neq\beta_H$, the $(\rho,\xi)$ space has the topology of a cone with a nonzero deficit angle $2\pi(1-\frac{\beta}{\beta_H})$. It should be noted that $r=r_h$ corresponds to $\rho=0$, which is the vertex of the cone. Thus, the scalar curvature $R$ diverges at the horizon and the standard formulas of the Riemannian geometry are not applicable here.

Regarding the part of the electromagnetic field, if we choose the electric potential as $\Phi_E\neq\Phi_{EH}$, the electromagnetic gauge potential in Eq.~(\ref{eq:999}) will be rewritten as
\be
\ba
\label{eq:20}
A_{\mu}=-(\frac{q_E}{r}-\Phi_E)dt+q_M(1-\cos\theta)d\phi.
\ea
\ee
After a simple calculation of $\sqrt{|g^{tt}A_tA_t|}$, we can find that $A_t$ also diverges at the horizon.

Because the divergences of $R$ and $A_{\mu}$ are located at the horizon $\Sigma_-$, the off-shell corrections to the Euclidean action are actually $I_{\Sigma_-}$ mentioned in the previous section. Before dealing with the two kinds of divergences at the horizon, we impose the gravitational Hamiltonian constraint $G^{\tau}_{\tau}+\Lambda g^{\tau}_{\tau}=8\pi T^{\tau}_{\tau}$ to eliminate the cosmological constant $\Lambda$ in the Euclidean action. Such a procedure has been used in Ref.~\cite{CK} to derive the reduced action of the charged AdS black hole. In our case, the Euclidean action is now rewritten as
\be
\ba
\label{eq:21}
I'_{bulk}=-\frac{1}{16\pi}\int_M d^4x\sqrt{g}(R+2G^{\tau}_{\tau}-16\pi T^{\tau}_{\tau}-F^2).
\ea
\ee

During the calculations of the off-shell corrections by the scheme of regularization, we have both used the gravitational Hamiltonian constraint and the electromagnetic Gauss's-law constraint~\cite{CJ, CK}. As said in~\cite{CJ, CK}, the solutions of the two constraints depend on two free parameters $r_h$ and $q_E$, which represent two physical degrees of freedom. By inserting the solutions of the two constraints into the Einstein-Maxwell action, one can obtain the reduced action. However, in~\cite{CJ, CK}, the black hole is enclosed in a cavity with a finite radius $r_B$, and the specified temperature and electric potential are determined by an observer at rest at $r_B$. Additionally, the manifold is always regular in~\cite{CJ, CK}.

\subsection{The contribution of $T\neq T_H$}
In this subsection, we will calculate the off-shell correction to the Euclidean action when $T\neq T_H$. Due to the presence of the conical singularity, it is necessary to regularize the divergence stemming from this singularity. Our procedure is shown as follows.

At first, we rewrite Eq.~(\ref{eq:17}) and Eq.~(\ref{eq:19}) as 
\be
\ba
\label{eq:22}
r(\rho)=r_h+\frac{f'(r_h)}{4}\rho^2
\ea
\ee
and
\be
\ba
\label{eq:23}
ds^2=\frac{\rho^2\beta^2}{\beta_H^2}d\psi^2+d\rho^2+r^2(\rho)d\theta^2+r^2(\rho)\sin^2\theta d\phi^2,
\ea
\ee
where $\psi=\frac{2\pi\tau}{\beta}$ has a period of $2\pi$.

The objective is to smooth out the conical deficit by replacing $\frac{\rho^2\beta^2}{\beta_H^2}$ with a regular function $a^2(\rho)$, which fulfills $a'(0)=1$ and $a'(\epsilon)=\frac{\beta}{\beta_H}$~\cite{CA, CO}. It means that the topology of $(\rho,\psi)$ space is now a disk at the horizon but a cone at $\rho=\epsilon$, so a limit $\epsilon\to 0$ should be taken to recover the conical singularity finally. The metric in Eq.~(\ref{eq:23}) is then rewritten as
\be
\ba
\label{eq:24}
ds^2=a^2(\rho)d\psi^2+d\rho^2+r^2(\rho)d\theta^2+r^2(\rho)\sin^2\theta d\phi^2.
\ea
\ee

In the vicinity of the horizon, we now use this metric to calculate the scalar curvature $R$ and the component of the Einstein tensor $G^{\psi}_{\psi}$ as
\be
\ba
\label{eq:25}
R=-\frac{2a''(\rho)}{a(\rho)}+\frac{2}{r_h^2}-\frac{8\pi}{\beta_Hr_h}
\ea
\ee
and 
\be
\ba
\label{eq:26}
G^{\psi}_{\psi}=-\frac{1}{r_h^2}+\frac{4\pi}{\beta_Hr_h},
\ea
\ee
where we have used $r(\epsilon)=r_h$ in the limit of $\epsilon\to 0$, and the vicinity of horizon represents that $0<\rho<\epsilon$ with $\epsilon\to 0$. Furthermore, we should note that $y''(x)$ represents the derivative of $y$ with respect to $x$. For example, $f'(r_h)$ in Eq.~(\ref{eq:22}) is the derivative of $f(r)$ with respect to $r$ at $r=r_h$. $a''(\rho)$ in Eq.~(\ref{eq:25}) is the second derivative of $a(\rho)$ with respect to $\rho$.

Substituting Eq.~(\ref{eq:25}) and Eq.~(\ref{eq:26}) into the gravitational part of the action in Eq.~(\ref{eq:21}), the correction to the Euclidean action resulting from the conical singularity can be calculated as 
\be
\ba
\label{eq:27}
I_{cc}&=\lim_{\epsilon\to 0}-\frac{1}{16\pi}\int_0^{\epsilon}d{\rho}\int_0^{2\pi}d\psi\int_0^{\pi}d\theta\int_0^{2\pi}d\phi \sqrt{g}(R+2G_{\psi}^{\psi})\\
&=\lim_{\epsilon\to 0}-\frac{1}{16\pi}\int_0^{\epsilon}d{\rho}\int_0^{2\pi}d\psi\int_0^{\pi}d\theta\int_0^{2\pi}d\phi[-2r_h^2a''(\rho)\sin\theta]\\
&=-\pi r_h^2(1-\frac{\beta}{\beta_H}).
\ea
\ee
We can find that the correction to the Euclidean action is independent of the choice of the regular function $a(\rho)$.


\subsection{The contribution of $\Phi_E\neq \Phi_{EH}$}
In this subsection, we will calculate the off-shell correction to the Euclidean action when $\Phi_E\neq\Phi_{EH}$. Since only the component $A_t$ of the electromagnetic gauge potential diverges at the horizon, we will only consider the electric part of the electromagnetic gauge potential here. After we use the analytic continuation $t\to i\tau$ in Eq.~(\ref{eq:20}), the electromagnetic gauge potential can be rewritten as 
\be
\ba
\label{eq:28}
A_{\mu}&=-i[\frac{q_E}{r(\rho)}-\Phi_E]d\tau+q_M(1-\cos\theta)d\phi\\
&=-\frac{i\beta}{2\pi}[\frac{q_E}{r(\rho)}-\Phi_E]d\psi+q_M(1-\cos\theta)d\phi,
\ea
\ee
where $\psi=\frac{2\pi\tau}{\beta}$ has a period of $2\pi$ and $A_{\psi}=-\frac{i\beta}{2\pi}[\frac{q_E}{r(\rho)}-\Phi_E]$ is the part that we care about.

Then, the nonvanishing components of the Maxwell field strength tensor $F_{\mu\nu}$ and $T^{\psi}_{\psi}$ can be calculated as
\be
\ba
\label{eq:29}
F_{\rho\psi}=-F_{\psi\rho}=A_{\psi}'(\rho),\quad F^{\rho\psi}=-F^{\psi\rho}=\frac{\beta_H^2A_{\psi}'(\rho)}{\beta^2\rho^2},
\ea
\ee
and 
\be
\ba
\label{eq:30}
T^{\psi}_{\psi}&=g_{\psi\mu}T^{\psi\mu}=\frac{F_{\psi\rho}F^{\psi\rho}}{8\pi},
\ea
\ee
where $T^{\mu\nu}=\frac{1}{4\pi}[F^{\mu\sigma}F^{\nu}_{\sigma}-\frac{1}{4}g^{\mu\nu}F_{\sigma k}F^{\sigma k}]$ is the Maxwell stress-energy tensor.

Substituting Eq.~(\ref{eq:29}) and Eq.~(\ref{eq:30}) into the electromagnetic part of the Euclidean action in Eq.~(\ref{eq:21}), we can write the off-shell correction to the Euclidean action resulting from the divergence of $A_{\mu}$ as
\be
\ba
\label{eq:31}
I_{ce}&=\lim_{\epsilon\to 0}-\frac{1}{16\pi}\int_0^{\epsilon}d\rho\int_0^{2\pi}d\psi\int_0^{\pi}d\theta\int_0^{2\pi}d\phi\sqrt{g}(-16\pi T^{\psi}_{\psi}-2F_{\rho\psi}F^{\rho\psi})\\
&=\lim_{\epsilon\to 0}\int_0^{\epsilon}d\rho[\frac{2\pi\beta_H r^2(\rho)A_{\psi}'^2(\rho)}{\beta\rho}].
\ea
\ee

Then, we impose the Gauss's-law constraint as done in the procedure of calculating the reduced action in~\cite{CJ, CK}. This is actually the nontrivial Maxwell equation as
\be
\ba
\label{eq:32}
\frac{d}{d\rho}[\frac{\beta_H r^2(\rho)A_{\psi}'(\rho)}{\beta\rho}]=0,
\ea
\ee
or equivalently,
\be
\ba
\label{eq:33}
\frac{\beta_H r^2(\rho)A_{\psi}'(\rho)}{\beta\rho}=constant.
\ea
\ee
If we substitute Eq.~(\ref{eq:6}), Eq.~(\ref{eq:22}), and Eq.~(\ref{eq:28}) into Eq.~(\ref{eq:33}), the constant can be calculated as $iq_E$.

By substituting Eq.~(\ref{eq:33}) into Eq.~(\ref{eq:31}) once, $I_{ce}$ can be rewritten as
\be
\ba
\label{eq:34}
I_{ce}=\lim_{\epsilon\to 0}2\pi iq_E\int_0^{\epsilon}d\rho A_{\psi}'(\rho).
\ea
\ee

Now, we will use a regularization scheme that is similar to the procedure used to handle the conical singularity. Specifically, we replace $A_{\psi}$ with a function $b(\rho)$ such that $b(0)=-\frac{i\beta}{2\pi}[\frac{q_E}{r(0)}-\Phi_{EH}]=0$ and $b(\epsilon)=-\frac{i\beta}{2\pi}[\frac{q_E}{r(\epsilon)}-\Phi_{E}]$. Different from $A_{\psi}$, the function $b(\rho)$ does not diverge at the horizon, and we finally take the limit $\epsilon\to 0$ to recover the singularity. However, the derivation of $b(\rho)$ diverges near the horizon as
\be
\ba
\label{eq:36}
b'(\epsilon)&=\lim_{\epsilon\to 0}\frac{b(\epsilon)-b(0)}{\epsilon}\\
&=\lim_{\epsilon\to 0}\frac{i\beta (\Phi_E-\Phi_{EH})}{2\pi\epsilon}.
\ea
\ee

Then, Eq.~(\ref{eq:34}) can be rewritten as 
\be
\ba
\label{eq:35}
I_{ce}&=\lim_{\epsilon\to 0}2\pi iq_E[b(\epsilon)-b(0)]\\
&=\beta q_E(\Phi_{EH}-\Phi_E),
\ea
\ee
where we have used $r(\epsilon)=r_h$ in the limit of $\epsilon\to 0$, and the result of $I_{ce}$ is independent of the choice for $b(\rho)$.

One may note that if we use Eq.~(\ref{eq:33}) twice in Eq.~(\ref{eq:31}), we will obtain
\be
\ba
\label{eq:6313}
I_{ce}&=\lim_{\epsilon\to 0}\int_0^{\epsilon}d\rho[\frac{2\pi\beta_H r^2(\rho)A_{\psi}'^2(\rho)}{\beta\rho}]\\
&=\lim_{\epsilon\to 0}\int_0^{\epsilon}d\rho [-\frac{2\pi\beta q_E^2\rho}{\beta_Hr^2(\rho)}]\\
&=0.
\ea
\ee
However, if we use the regularization scheme twice by replacing $A_{\psi}'^2(\rho)$ with $b'^2(\rho)$ in Eq.~(\ref{eq:31}), we will obtain 
\be
\ba
\label{eq:6314}
I_{ce}&=\lim_{\epsilon\to 0}\int_0^{\epsilon}d\rho[\frac{2\pi\beta_H r^2(\rho)b'^2(\rho)}{\beta\rho}]=\infty,
\ea
\ee
because $\int_0^{\epsilon}d\rho b'(\rho)$ is finite and non-zero, but $\frac{b'(\rho)}{\rho}$ diverges when $\epsilon\to 0$. Recalling the procedure for dealing with the conical singularity. If we use the metric in Eq.~(\ref{eq:23}) instead of the metric in Eq.~(\ref{eq:24}), i.e., if we do not employ the regularization scheme, we find that the value of $I_{cc}$ in Eq.~(\ref{eq:27}) is also $0$. Consequently, the regularization is deemed too weak if Eq.~(\ref{eq:33}) is used twice, or too strong if $A'_{\psi}(\rho)$ is replaced by $b'(\rho)$ twice. Additionally, it can be proven that in order to obtain a finite and non-zero value for $I_{ce}$, we can only replace $A'_{\psi}(\rho)$ by $b'(\rho)$ once. Furthermore, if we substitute Eq.~(\ref{eq:33}) twice in Eq.~(\ref{eq:31}), the reduced action will be independent of the free parameter $\Phi_E$. This indicates that the physical degree of freedom $\Phi_E$ is lost.

In conclusion, the generalized free energy can be calculated as
\be
\ba
\label{eq:522}
G_{gen}&=\frac{I_{M/\Sigma_-}+I_{cc}+I_{ce}}{\beta}\\
&=(\frac{r_h}{2}+\frac{r_h^3}{2l^2}+\frac{q_E^2+q_M^2}{2r_h})-T\pi r_h^2-\Phi_E q_E\\
&=M-TS-\Phi_Eq_E,
\ea
\ee
and we have obtained the same generalized free energy as the result from the thermodynamic perspective in Eq.~(\ref{eq:12}). For the purpose of simplification, we will use the symbols $r$ and $Q$ to represent $r_h$ and $q_E$ in the following paper. In Appendix~\ref{Apa}, we present another possible approach for calculating the off-shell corrections to the Euclidean action and its associated generalized free energy. This approach includes adding a boundary near the horizon to exclude the singularities. Finally, both methods yield the same result.

\begin{figure}[t]
\centering
\includegraphics[width=0.96\textwidth]{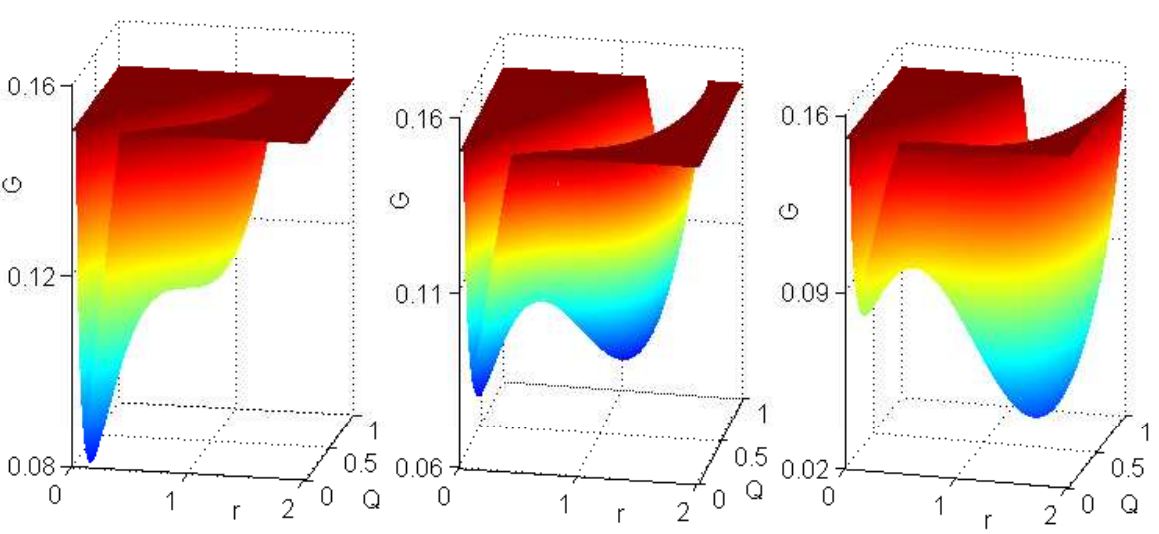}
\caption{The generalized free energy landscape is plotted at $P=0.042$, $\Phi_E=0.5$ and $q_M=0.1$. We have truncated the free energy landscape at $G=0.16$, and the portion where $G>0.16$ is represented by $G=0.16$ in the figure. The temperatures of the three subfigures are different, and they are $T=0.141$, $0.15$ and $0.16$ from left to right. When $T=0.15$, the two minima have equal depth in the well.}
\label{Fig1}
\end{figure}

In Fig.~\ref{Fig1}, we have plotted the generalized free energy landscape of the dyonic AdS black hole as a function of $r$ and $Q$ in $P=0.042$, $\Phi_E=0.5$, and $q_M=0.1$. There are two locally stable states corresponding to the minima of the free energy landscape, and one unstable state corresponding to the saddle point of the free energy landscape. These three states are on-shell, while all other states on the landscape are off-shell. We refer to these three on-shell black holes by their sizes as small, intermediate, and large black holes, with radii denoted as $r_s$, $r_m$, and $r_l$, respectively. When $T=0.15$, the free energies of the small and large black holes are equal, and they are both globally stable states. When $T<0.15$, the small black hole state is the globally stable state with a lower free energy. When $T>0.15$, the large black hole state is the globally stable state with a lower free energy.


\section{The dynamical state transition for the dyonic AdS black hole in the grand canonical ensemble}
In the previous section, we have derived the generalized free energy of the dyonic AdS black hole in the grand canonical ensemble by the gravitational path integral approach. In the free energy landscape at a fixed temperature, as shown in Fig.~\ref{Fig1}, it seems impossible for a locally stable state to overcome the barrier of the unstable state and transform into another locally stable state. However, if we consider the black hole as a thermal entity, the thermal fluctuations are unavoidable and will induce the state transition between these two locally stable states. In analogy to the motion of a Brownian particle, the dynamics of state transition for the dyonic AdS black hole in the grand canonical ensemble are described by the Langevin equation that governs the stochastic evolution of the order parameters $r$ and $Q$ as
\be
\ba
\label{eq:40}
\frac{d^2\vec{x}}{dt^2}=-\tilde{\gamma}\frac{d\vec{x}}{dt}-\nabla G(\vec{x})+\vec{\xi}(\vec{x},t),
\ea
\ee
where $\vec{x}=(r,Q)$ and $\tilde{\gamma}$ is the two-dimensional friction coefficient matrix. On the left side, we have the inertia term. On the right side, the first force is the friction opposing the direction of motion in the order parameter space, the second force is the driving force resulting from the free energy landscape, and the third force is the fluctuating stochastic force. For simplicity, we make the following assumptions: (1) The friction coefficient matrix $\tilde{\gamma}$ is isotropic and homogeneous, so $\tilde{\gamma}_{ij}=\gamma\delta_{ij}$ and $\gamma$ is a constant. (2) $\vec{\xi}(\vec{x},t)$ is Gaussian white noise and the dynamics of the state transition are Markovian. (3) $\gamma$ is assumed to be very large, so $\frac{d\vec{x}}{dt}$ and $\vec{x}$ have significantly different characteristic time scales (See Appendix~\ref{Asef} for an illustrative example). Specifically, $\frac{d\vec{x}}{dt}$ is the fast variable that rapidly converges to a quasi-stable value compared to the slow variable $\vec{x}$. In the case of a stochastic system, the behavior on a very short time scale is often not of interest. A prime example is the Brownian particle, where only the position is observed, while the momentum is not of interest. Consequently, the adiabatic approximation allows us to neglect the inertia term $\frac{d^2\vec{x}}{dt^2}$ in Eq.~(\ref{eq:40}), as $\frac{d\vec{x}}{dt}$ quickly reaches quasi-stability with $\frac{d^2\vec{x}}{dt^2}\to 0$~\cite{QAA, QAB, QAC, CR}. Then, Eq.~(\ref{eq:40}) can be rewritten as the overdamped Langevin equation
\be
\ba
\label{eq:41}
&\frac{dr}{dt}=-\frac{\partial G(r,Q)}{\gamma\partial r}+\eta_1(r,Q,t),\\
&\frac{dQ}{dt}=-\frac{\partial G(r,Q)}{\gamma\partial Q}+\eta_2(r,Q,t).
\ea
\ee
$\eta_i(r,Q,t)$ are the Gaussian white noises, which satisfy 
\be
\ba
\label{eq:42}
<\eta_i(r,Q,t)>=0,\quad <\eta_i(r,Q,t)\eta_j(r,Q,0)>=2D\delta(t)\delta_{ij},\quad i,j=1,2.
\ea
\ee
$D$ is the diffusion coefficient associated with the friction coefficient $\gamma$ by the fluctuation-dissipation theorem
\be
\ba
\label{eq:43}
D\gamma=k_BT,
\ea
\ee
which states that the friction is actually determined by the correlation of the fluctuating stochastic force. 

We can now formulate the stochastic dynamics described by the Langevin equation with the Onsager-Machlup functional, and the probability from the initial state $(r_i, Q_i)$ at time $t_0=0$ to the final state $(r_f, Q_f)$ at time $t$ can be quantified as~\cite{CQ, CR}
\be
\ba
\label{eq:44}
P(r_f,Q_f,t;r_i,Q_i,t_0)&=\int D\vec{x} \exp\{-\int L dt\}\\
&=\int D\vec{x}\exp\{-\int[\frac{(\frac{dr}{dt}-f_r)^2}{4D}+\frac{(\frac{dQ}{dt}-f_Q)^2}{4D}+\frac{\partial_rf_r+\partial_Qf_Q}{2}]dt\},
\ea
\ee
where $(f_r,f_Q)=(-\beta D\partial_r G, -\beta D\partial_Q G)$, $L$ is the stochastic Lagrangian and $D\vec{x}$ represents the sum of all the paths connecting the initial state and the final state. Because the friction coefficient $\gamma$ is very large in our case, the corresponding diffusion coefficient $D$ is very small and the last term in stochastic Lagrangian can be ignored as
\be
\ba
\label{eq:45}
L=\frac{(\frac{dr}{dt}-f_r)^2}{4D}+\frac{(\frac{dQ}{dt}-f_Q)^2}{4D}.
\ea
\ee

From Eq.~(\ref{eq:44}), we can see that the various paths contribute to different weights, which are on the exponentials. This indicates that the dominant path has the largest weight, which is significantly larger than the weights of other paths. Thus, we can just consider the contribution of the dominant path, which satisfies the Euler-Lagrange equation from the maximization of the weights or minimization of the action as
\be
\ba
\label{eq:46}
&\frac{d}{dt}\frac{\partial L}{\partial\dot{r}}-\frac{\partial{L}}{\partial r}=0,\\
&\frac{d}{dt}\frac{\partial L}{\partial\dot{Q}}-\frac{\partial{L}}{\partial Q}=0.
\ea
\ee

Substituting Eq.~(\ref{eq:45}) into Eq.~(\ref{eq:46}) and integrating them, we can obtain
\be
\ba
\label{eq:47}
\frac{1}{4D}\dot{r}^2+\frac{1}{4D}\dot{Q}^2-(\frac{1}{4D}f_r^2+\frac{1}{4D}f_Q^2)=E,
\ea
\ee
where $E$ is an integration constant. Eq.~(\ref{eq:47}) can be considered as an energy conservation equation, where $\frac{1}{4D}\dot{r}^2+\frac{1}{4D}\dot{Q}^2$ is the kinetic energy term, $V(r,Q)=-(\frac{1}{4D}f_r^2+\frac{1}{4D}f_Q^2)$ is the effective potential, and $E$ is the total energy. Thus, the stochastic dynamics can be regarded as the dynamics of a particle with mass $\frac{1}{2D}$ moving in the two-dimensional potential $V(r, Q)$. 

In principle, we can solve Eq.~(\ref{eq:46}) to obtain the dominant path. However, it is not an easy task for such two-dimensional differential equations with boundary-value conditions rather than initial-value conditions. Fortunately, the dynamics are energy-conserving and time-reversible, so we can switch the dynamics from the time-dependent Newtonian description to the energy-dependent Hamilton-Jacobi description~\cite{CS, CT, CU}. In other words, we will not solve the time-dependent trajectories $r(t)$ and $Q(t)$, but rather solve the time-independent path $Q(r)$. The Lagrangian $L$ is associated with the Hamiltonian $H$ by the Legendre transformation as
\be
\ba
\label{eq:10111}
L=\sum_k p_k\dot{q_k}-H, \quad k=1,2,
\ea
\ee
where $(q_1,q_2)=(r,Q)$ and $p_k$ is given by
\be
\ba
\label{10112}
p_k=\frac{\partial L}{\partial \dot{q_k}}=\frac{\dot{q_k}-f_{q_k}}{2D}.
\ea
\ee

Substituting Eq.~(\ref{eq:10111}) and the energy conservation equation~(\ref{eq:47}) into the effective action $S=\int L dt$, we can obtain
\be
\ba
\label{eq:48}
S_{HJ}=\int_{\vec{x_i}}^{\vec{x_f}}\sqrt{\frac{E+\frac{1}{4D}f_r^2+\frac{1}{4D}f_Q^2}{D}}dl,
\ea
\ee
where $dl=\sqrt{dr^2+dQ^2}$ is an infinitesimal displacement along the path and we have abandoned the boundary terms because they do not play a role in the calculations of the dominant path with minimal action. In our case, we take $E$ as $E=V(\vec{x_i})=V(\vec{x_f})=0$, which corresponds to the longest transition time in Eq.~(\ref{eq:47}). Without loss of generality, we set $D=1$ throughout the paper.

\begin{figure}[t]
\centering
\includegraphics[width=0.48\textwidth]{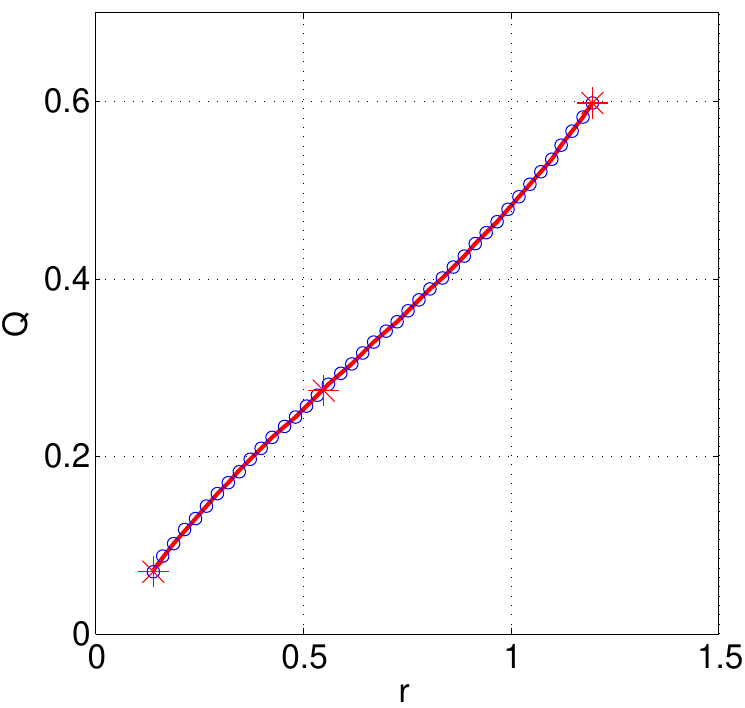}
\caption{The blue circles are the dominant path in the order parameter space obtained by the simulated annealing and conjugate gradient algorithm at $P=0.042$, $T=0.15$, $\Phi_E=0.5$, and $q_M=0.1$. The three points marked by the red stars are the on-shell black holes with $(r, Q)=(0.140,0.070),(0.549,0.275),(1.198,0.599)$. The red line represents the polynomial $Q=h(r)$ in Eq.~(\ref{eq:51}).}
\label{Fig2}
\end{figure}

The dominant path $Q(r)$ can be obtained by minimizing the discretized action
\be
\ba
\label{eq:49}
S_{HJ}=\sum_{n=0}^{N-1}\{\sqrt{\frac{E+\frac{1}{4D}f_r^2+\frac{1}{4D}f_Q^2}{D}}\Delta l_{n,n+1}+\lambda (\Delta l_{n,n+1}-<\Delta l>)^2\},
\ea
\ee
where
\be
\ba
\label{eq:50}
(\Delta l)^2_{n,n+1}=[r(n+1)-r(n)]^2+[Q(n+1)-Q(n)]^2.
\ea
\ee
The last term in Eq.~(\ref{eq:49}) is a numerical technique used to introduce a penalty function, which keeps all the length elements close to their average and becomes irrelevant in the continuum limit~\cite{CS, CT, CU}. Then, we apply the simulated annealing and conjugate gradient algorithm to iterate and obtain the dominant path $Q(r)$ with minimal action. When $P=0.042$, $T=0.15$, $\Phi_E=0.5$, and $q_M=0.1$, we plot the dominant path in Fig.~\ref{Fig2} as the blue circles. Actually, the dominant path we obtain is described by a set of discrete points. However, we can approximate these points using a polynomial, and a suitable choice is the polynomial given by
\be
\ba
\label{eq:51}
Q=h(r)=1.0587r^7-5.1588r^6+10.6026r^5\\
-11.8481r^4+7.9026r^3-3.2489r^2+1.2294r-0.0559.
\ea
\ee

In Fig.~\ref{Fig2}, we have plotted $Q=h(r)$ by red line, which matches well with the blue circles. We should note that the weight of the dominant path is significantly larger than the weights of other paths, and the state transition can be considered along $Q=h(r)$. Therefore, only $r$ is a free parameter and the two-dimensional dynamics $r(t)$ and $Q(t)$ can be transformed into the one-dimensional dynamics $r(t)$. We replace $Q$ in Eq.~(\ref{eq:45}) with $h(r)$ and solve the Euler-Lagrange equation again, it yields
\be
\ba
\label{eq:52}
\frac{1+h'^2(r)}{2D}\ddot{r}+\frac{h'(r)h''(r)}{2D}\dot{r}^2-\frac{f_r(r)}{2D}\frac{df_r(r)}{dr}-\frac{f_Q(r)}{2D}\frac{df_Q(r)}{dr}=0.
\ea
\ee

Integrating Eq.~(\ref{eq:52}), we can obtain the energy conservation equation as
\be
\ba
\label{eq:53}
\frac{1+h'^2(r)}{4D}\dot{r}^2-\frac{f_r^2(r)+f_Q^2(r)}{4D}=E,
\ea
\ee
where $E$ is set to zero corresponding to the longest transition time. Interestingly, the mass of the equivalent particle moving in one-dimensional effective potential is $r$-dependent as $\frac{1+h'^2(r)}{2D}$. 

\begin{figure}[t]
\centering
\includegraphics[width=0.48\textwidth]{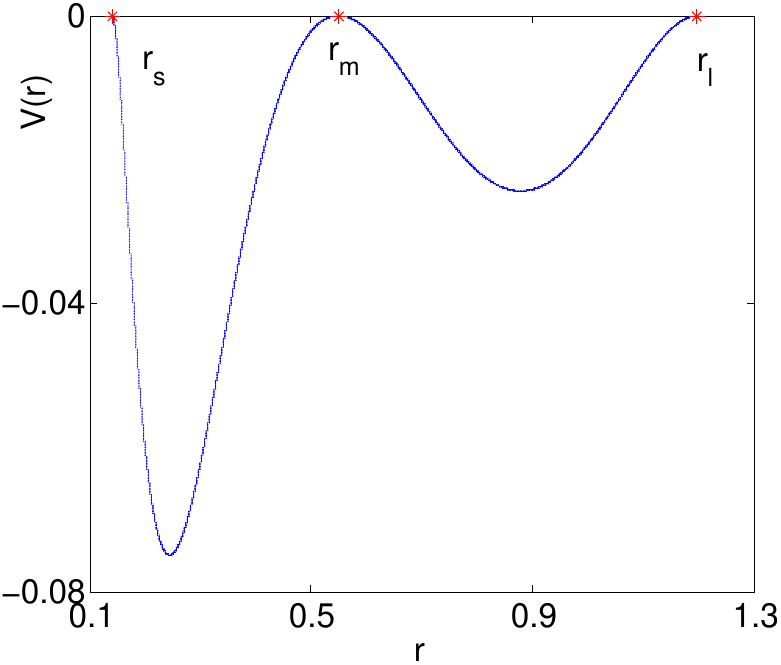}
\caption{The one-dimensional effective potential is plotted with a blue line at $P=0.042$, $T=0.15$, $\Phi_E=0.5$, and $q_M=0.1$. The three points marked by the red stars are the on-shell black holes, with radii $r_s$, $r_m$, and $r_l$ from left to right.}
\label{Fig3}
\end{figure}

\begin{figure}[t]
\centering
\includegraphics[width=0.65\textwidth]{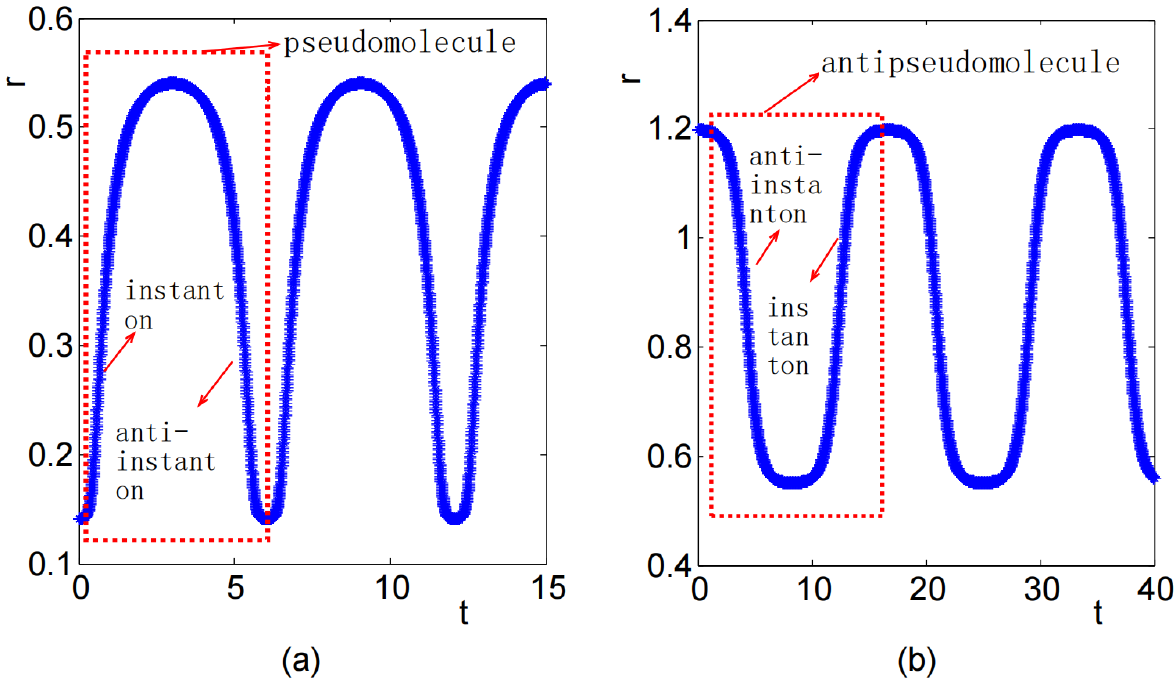}
\caption{Some sections of the time-dependent domain path $r(t)$ are shown at $P=0.042$, $T=0.15$, $\Phi_E=0.5$, and $q_M=0.1$. The left figure consists of many $a$ pseudomolecules, while the right figure consists of many $c$ pseudomolecules(or called antipseudomolecules). To demonstrate the formation of $b$ pseudomolecule, one can combine the instanton trajectory from $r_s$ to $r_m$ in the $a$ pseudomolecule with the instanton trajectory from $r_m$ to $r_l$ in the $c$ pseudomolecule. Conversely, the $d$ pseudomolecule is the time-reversed version of the $b$ pseudomolecule. The domain path is composed of a sequence of $a$, $b$, $c$, and $d$ pseudomolecules.}
\label{Fig4}
\end{figure}

In Fig.~\ref{Fig3}, we have plotted the one-dimensional effective potential $V(r)=-\frac{f_r^2(r)+f_Q^2(r)}{4D}$. The three on-shell black holes are located at the maxima where their effective potentials are equal to zero. The dynamics of state transition between the small and large black holes can be regarded as the dynamics of an effective particle moving between the left and right points marked by the red stars in Fig.~\ref{Fig3}. In the long time limit, the state transition between the small and large black holes can take place many times, indicating that the equivalent particle can move forth and back many times between the points $r_s$ and $r_l$. Consequently, the dominant path will consist of a series of the smallest units called pseudomolecules (or instanton and anti-instanton pairs shown in Fig.~\ref{Fig4}), with their initial and final states located at the locally stable states and the other states of the pseudomolecules being unstable. There are four kinds of pseudomolecules in total as
\be
\ba
\label{eq:54}
&a: r_s\to r_m\to r_s,
\quad b: r_s\to r_m\to r_l,\\
&c: r_l\to r_m\to r_l,
\quad d: r_l\to r_m\to r_s,
\ea
\ee
whose contributions to the probability are denoted as $w_a$, $w_b$, $w_c$, and $w_d$, respectively.

In Fig.~\ref{Fig4}, we have plotted some sections of the domain path $r(t)$ based on Eq.~(\ref{eq:52}). The dynamical trajectory of state transition is revealed by the domain path, which is composed of three on-shell states and many off-shell states. Therefore, it is necessary for us to introduce the off-shell states. From the path, it can be observed that there is no residence time for the off-shell states, indicating that they are unstable transient states. One might question why the unstable on-shell intermediate state with radius $r_m$ has a residence time. In fact, this is due to a numerical fault. When considering the stochastic Lagrangian in Eq.~(\ref{eq:44}), we have neglected the last term. Although this term is very small, it is not zero and results in the effective potential of $r_m$ being smaller than $r_s$ and $r_l$ in Fig.~\ref{Fig3}. Consequently, strictly speaking, there will be no residence time for the intermediate black hole.

Although the domain path has shown how the process proceeds during the state transition, a complete description of the dynamics should also encompass the rate that quantifies the time scale of the state transition between the small and large black holes. To calculate this rate, we initially calculate the probability $P(r_f,t;r_i,t_0)$ as follows. 

In our case, we choose the initial state as the small black hole state, and the final state can be either the small or large black hole state. We make the assumption that there are no interactions among the pseudomolecules, allowing us to calculate the probability $P(r_f,t;r_i,t_0)$ by separating it into the contribution of each part in the dilute gas approximation. The contribution of one pseudomolecule to the probability $P(r_f,t;r_i,t_0)$ is given by
\be
\ba
\label{eq:55}
W=\exp[-S]=\exp[-\int L dt],
\ea
\ee
with
\be
\ba
\label{eq:56}
L=\frac{1}{4D}\{[1+h'^2(r)]\dot{r}^2-2[f_r(r)+h'(r)f_Q(r)]\dot{r}+f_r^2(r)+f_Q^2(r)\}.
\ea
\ee

We should note that the downhill part of the free energy landscape along the trajectory of the pseudomolecule does not contribute to the probability. In other words, the Lagrangian is always zero during $r_m\to r_s$ and $r_m\to r_l$ of the pseudomolecule. This can be proven as follows. Firstly, we rewrite the energy conservation equation~(\ref{eq:53}) as
\be
\ba
\label{eq:57}
\dot{r}^2=\frac{f_r^2(r)+f_Q^2(r)}{1+h'^2(r)}.
\ea
\ee

Then, we substitute Eq.~(\ref{eq:57}) into Eq.~(\ref{eq:56}), and $L=0$ can be simplified as 
\be
\ba
\label{eq:58}
[f_r(r)+h'(r)f_Q(r)]\dot{r}=f_r^2(r)+f_Q^2(r).
\ea
\ee

\begin{figure}[t]
\centering
\includegraphics[width=0.48\textwidth]{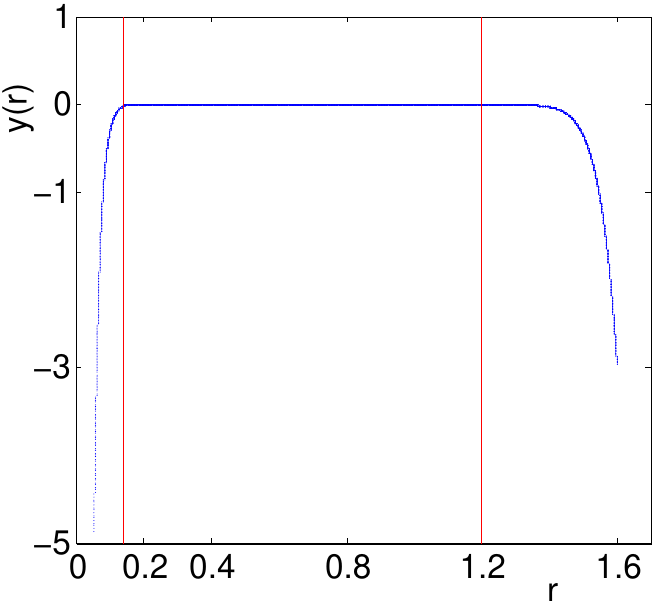}
\caption{$y(r)=f_Q(r)-h'(r)f_r(r)$ is plotted with a blue line at $P=0.042$, $T=0.15$, $\Phi_E=0.5$ and $q_M=0.1$. The left red line corresponds to $r=r_s$, and the right red line corresponds to $r=r_l$.}
\label{Fig5}
\end{figure}

If we square the two sides of Eq.~(\ref{eq:58}) and substitute Eq.~(\ref{eq:57}) into it, we can obtain 
\be
\ba
\label{eq:59}
f_Q(r)=h'(r)f_r(r).
\ea
\ee
We should note that Eq.~(\ref{eq:59}) gives $L=0$ or $L=[f_r^2(r)+f_Q^2(r)]/D$, which is determined by the sign of $[f_r(r)+h'(r)f_Q(r)]\dot{r}$ in Eq.~(\ref{eq:58}). Then, we introduce a function $y(r)=f_Q(r)-h'(r)f_r(r)$ and plot it in Fig.~\ref{Fig5}. From the figure, we can see that Eq.~(\ref{eq:59}) is always satisfied when $r$ is within the range of the radius of small black hole $r_s$ and the radius of large black hole $r_l$. However, the relation may be violated when $r$ is outside of this range. It is reasonable because the polynomial fitting of $Q$ by $h(r)$ in Eq.~(\ref{eq:51}) is only valid for $r_s\leq r\leq r_l$. If we analyze the sign of $\dot{r}$ and the free energy landscape, we can find that $L$ is always equal to $0$ during $r_m\to r_s$ and $r_m\to r_l$ in the trajectory of pseudomolecule. Therefore, the contributions to probability for four kinds of pseudomolecules satisfy
\be
\ba
\label{eq:60}
&w_a=-w_b=-W_1,\\
&w_c=-w_d=-W_2,
\ea
\ee
where $W_1$ and $W_2$ are given by Eq.~(\ref{eq:55}) whose integral domain is now taken as $t_s\to t_m$ and $t_l\to t_m$, respectively. The minus signs appearing in Eq.~(\ref{eq:60}) result from the presence of a turning point on the trajectory of the pseudomolecules, one can read Ref.~\cite{CV, CW} if interested in the origin. Eq.~(\ref{eq:60}) tells us that there are actually two independent categories of pseudomolecules, we classify $a$ and $b$ as the pseudomolecules and $c$ and $d$ as the antipseudomolecules.

Then, we choose the small black hole state as the final state and calculate the probability $P(r_s,t;r_s,t_0)$, which is a sum of all possible cases with the number of pseudomolecules from $0$ to infinity. When there is zero pseudomolecule, it means that the system always stays at the small black hole state for $t_0\sim t$. Eq.~(\ref{eq:44}) tells us that the residence time $t_0\sim t$ for the small black hole state will give a contribution to the probability $P(r_s,t;r_s,t_0)$ as $e^{V(r_s)(t-t_0)}$.

When there is one pseudomolecule, it can only be $a$ pseudomolecule, and the contribution is given by
\be
\ba
\label{eq:61}
\int_{t_0}^{\infty}dt_1e^{V(r_s)(t_1-t_0)}(-W_1)e^{V(r_s)(t-t_1)},
\ea
\ee
where $a$ pseudomolecule takes place at time $t_1$.

When there are two pseudomolecules, they can either be two $a$ pseudomolecules or $b\to d$ where the arrow represents the time sequence.
The contribution is given by
\be
\ba
\label{eq:62}
&\int_{t_0}^{\infty}dt_1\int_{t_1}^{\infty}dt_2\{e^{V(r_s)(t_1-t_0)}(-W_1)e^{V(r_s)(t_2-t_1)}(-W_1)e^{V(r_s)(t-t_2)}\}\\
&+\int_{t_0}^{\infty}dt_1\int_{t_1}^{\infty}dt_2\{e^{V(r_s)(t_1-t_0)}(W_1)e^{V(r_l)(t_2-t_1)}(W_2)e^{V(r_s)(t-t_2)}\},
\ea
\ee
where two pseudomolecules take place at time $t_1$ and $t_2$. The first and second terms in the equation correspond to $a\to a$ and $b\to d$, respectively.

When there are three pseudomolecules, they can be three $a$ pseudomolecules, $a\to b\to d$, $b\to d\to a$ or $b\to c\to d$. We continue such procedures and use $V(r_s)=V(r_l)=V$, the total probability $P(r_s,t;r_s,t_0)$ can be calculated by the sum of all possible cases with the number of pseudomolecules from $0$ to infinity as
\be
\ba
\label{eq:63}
P(r_s,t;r_s,t_0)&=e^{V(t-t_0)}-W_1\int_{t_0}^{\infty}dt_1e^{V(t_1-t_0)}e^{V(t-t_1)}\\
&+W_1(W_1+W_2)\int_{t_0}^{\infty}dt_1\int_{t_1}^{\infty}dt_2e^{V(t_1-t_0)}e^{V(t_2-t_1)}e^{V(t-t_2)}\\
&-W_1(W_1+W_2)^2\int_{t_0}^{\infty}dt_1\int_{t_1}^{\infty}dt_2\int_{t_2}^{\infty}dt_3e^{V(t_1-t_0)}e^{V(t_2-t_1)}e^{V(t_3-t_2)}e^{V(t-t_3)}+...\\
&=e^{V(t-t_0)}+W_1\sum_{n=1}^{\infty}(-1)^n(W_1+W_2)^{n-1}\int_{t_0}^{\infty}dt_1...\int_{t_{n-1}}^{\infty}dt_ne^{V(t_1-t_0)}...e^{V(t-t_n)}.
\ea
\ee
By using the Laplace transform, we obtain
\be
\ba
\label{eq:64}
P(s)=\frac{1}{s-V}-\frac{W_1}{W_1+W_2}[\frac{1}{s-V}-\frac{1}{s-V+W_1+W_2}],
\ea
\ee
where the effective potential $V$ is equal to 0.

Inverting the Laplace transform, we can obtain
\be
\ba
\label{eq:65}
P(r_s,t;r_s,0)=\frac{1}{W_1+W_2}[W_2+W_1e^{-(W_1+W_2)t}].
\ea
\ee
A similar procedure can be applied to the calculation of $P(r_l,t;r_s,0)$, which yields
\be
\ba
\label{eq:66}
P(r_l,t;r_s,0)=\frac{1}{W_1+W_2}[W_1-W_1e^{-(W_1+W_2)t}].
\ea
\ee

\begin{figure}[t]
\centering
\includegraphics[width=0.9\textwidth]{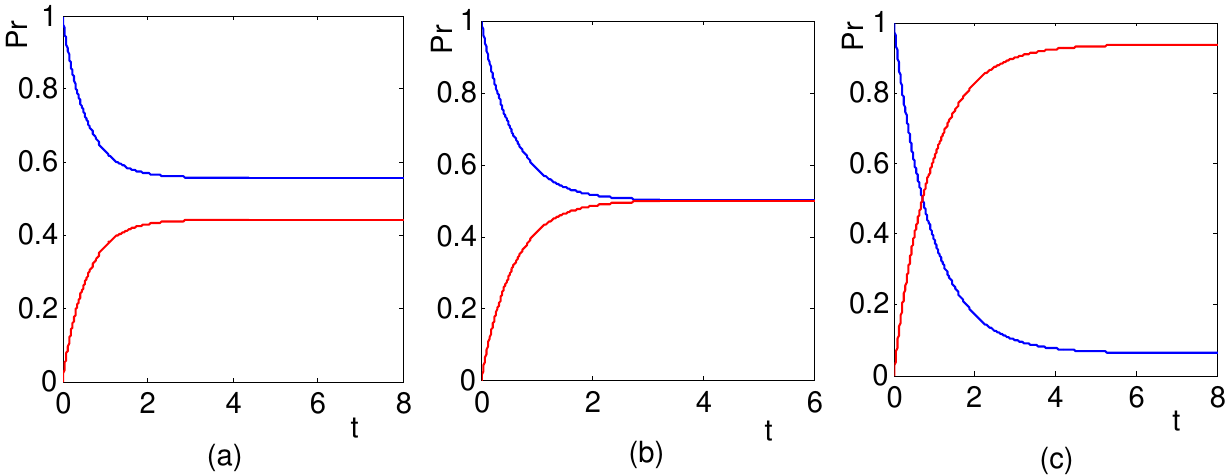}
\caption{The time dependencies of the probabilities are plotted at $P=0.042$, $\Phi_E=0.5$, and $q_M=0.1$. The blue lines represent $P(r_s,t;r_s,0)$, while the red lines represent $P(r_l,t;r_s,0)$. The three subfigures have different temperatures: $(a)$ $T=0.141$, $(b)$ $T=0.15$, and $(c)$ $T=0.21$.}
\label{Fig6}
\end{figure}

\begin{figure}[t]
\centering
\includegraphics[width=0.48\textwidth]{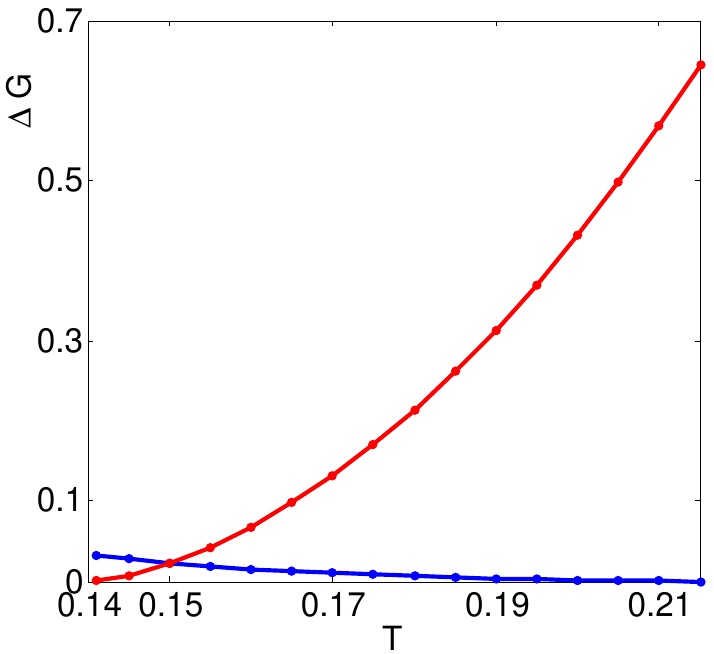}
\caption{The temperature dependencies of the barrier heights in the free energy landscape are plotted at $P=0.042$, $\Phi_E=0.5$, and $q_M=0.1$. The blue line represents the barrier height $G_m-G_s$ between the intermediate and the small black hole states, while the red line represents the barrier height $G_m-G_l$ between the intermediate and the large black hole states. Except for the leftmost point with $T=0.141$, the temperature interval between other points is $0.005$. If the temperature goes beyond the range of the curve, there will not be three on-shell states, and the phase transition will no longer occur. Actually, this can be observed from the barrier heights, as they gradually approach zero at $T=0.141$ and $0.215$. This indicates that one on-shell black hole state will disappear.}
\label{Fig7}
\end{figure}

In Fig.~\ref{Fig6}, we have plotted the time dependencies of $P(r_s,t;r_s,0)$ and $P(r_l,t;r_s,0)$ at different temperatures. Due to thermal fluctuations, both the small and large black hole states can occur with different steady probabilities. The steady probability is governed by the free energy according to the Boltzmann distribution, and it can indicate thermodynamic stability. In Fig.~\ref{Fig7}, we have plotted the barrier heights $G_m-G_s$ and $G_m-G_l$ in the free energy landscape. When $T=0.15$, the small and large black holes have the same free energy, and their steady probabilities are both equal to $0.5$ as shown in Fig.~\ref{Fig6}. When $T<0.15$, the steady probability of the small black hole is greater than that of the large black hole due to its lower free energy, making the small black hole globally stable. When $T>0.15$, the steady probability of the small black hole is smaller than that of the large hole due to its higher free energy, making the large black hole globally stable.

As we know, the master equation is a powerful tool to describe the time evolution of the probability distribution for the Markov process. In our system, there are two locally stable states, and the classical master equation can be written as
\be
\ba
\label{eq:67}
\frac{dP(r_s,t;r_s,0)}{dt}=-k_{1} P(r_s,t;r_s,0)+k_{2}P(r_l,t;r_s,0),
\ea
\ee
where $k_{1}$ is the transition rate from the small to large black hole state, and $k_{2}$ is the transition rate from the large to small black hole state. Combining Eq.~(\ref{eq:67}) with the probability conserved equation $P(r_s,t;r_s,0)+P(r_l,t;r_s,0)=1$, we can obtain
\be
\ba
\label{eq:68}
P(r_s,t;r_s,0)=\frac{1}{k_{1}+k_{2}}[k_{2}+k_{1}e^{-(k_{1}+k_{2})t}],\\
P(r_l,t;r_s,0)=\frac{1}{k_{1}+k_{1}}[k_{1}-k_{1}e^{-(k_{1}+k_{1})t}].
\ea
\ee

Comparing Eq.~(\ref{eq:68}) with Eq.~(\ref{eq:65}) and Eq.~(\ref{eq:66}), we can easily see that the contributions $W_1$ of pseudomolecule and $W_2$ of antipseudomolecule to the probability are actually the transition rates from the small to large and large to small black hole states, respectively. That is, $k_1=W_1$ and $k_2=W_2$.

\begin{figure}[t]
\centering
\includegraphics[width=0.9\textwidth]{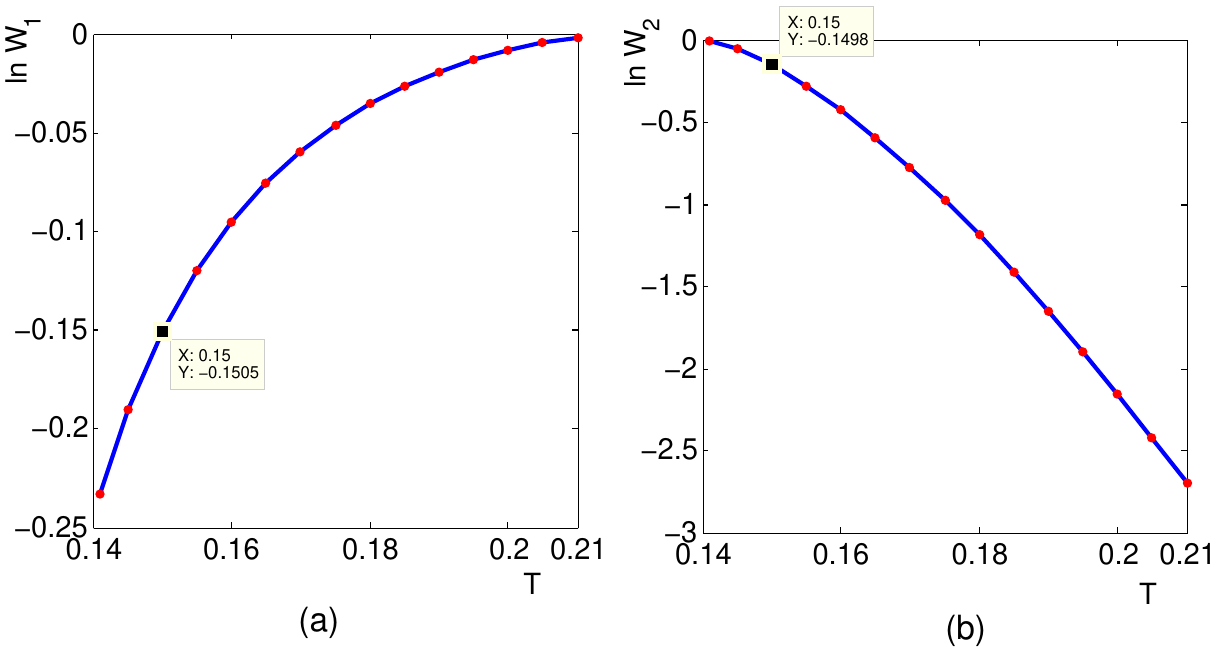}
\caption{The kinetic rates of the state transitions between the small and large black hole states are plotted at $P=0.042$, $\Phi_E=0.5$, and $q_M=0.1$. Figure $(a)$ on the left represents the transition rate from the small to large black holes, while figure $(b)$ on the right represents the transition rate from the large to small black holes. The horizontal axis represents temperature, and the vertical coordinate represents the logarithm of the transition rate. Except for the leftmost point with $T=0.141$, the temperature interval between other points is $0.005$.}
\label{Fig8}
\end{figure}

In Fig.~\ref{Fig8}, we have plotted the kinetic rates of state transitions between the small and large black hole states. It should be noted that although the steady probabilities of the small and large black hole states are determined by the values of their free energies, the transition rates between them are determined by the barrier heights in the free energy landscape. As the temperature increases, the transition rate from the small to large black hole state increases, while the transition rate from the large to small black hole state decreases. This is consistent with the change in the barrier height of the free energy with temperature shown in Fig.~\ref{Fig7}, where $G_m-G_s$ decreases and $G_m-G_l$ increases as the temperature increases. This indicates that the small black hole state needs to overcome a lower barrier height to switch to the large black hole state, resulting in an increased transition rate. On the other hand, the large black hole state needs to overcome a higher barrier height to switch to the small black hole state, resulting in a decreased transition rate. Additionally, when $T=0.15$, both the small and large black hole states are globally stable with equal free energy basin depth, and the transition rates $W_1$ and $W_2$ are equal. For $T<0.15$, the transition rate from the large to small black hole state is larger than that from the small to large black hole state, indicating that the system is finally dominated by the small black hole state. For $T>0.15$, the transition rate from the small to large black hole state is larger than that from the large to small black hole state, indicating that the system is finally dominated by the large black hole state. If we gradually increase the temperature, the system will dynamically switch from being finally dominated by the small black hole state to being finally dominated by the large black hole state. This is a signature of the liquid-gas type phase transition for the dyonic AdS black hole in the phase diagram.


\section{Conclusion and Discussion}
\label{CaDi}
In this paper, we have studied the generalized free energy and the dynamical state transition of the dyonic AdS black hole in the grand canonical ensemble. Considering the dyonic AdS black hole as a system in contact with a thermal and particle bath located at infinity, the temperature and the electric potential of the ensemble are external adjustable parameters. However, when we adjust them, the scalar curvature $R$ and the electromagnetic gauge potential $A_{\mu}$ are usually divergent. We provide two methods to deal with the divergences, and the reduced action is calculated to be the same for both methods. For the first method, we regularize them and incorporate the off-shell corrections into the Einstein-Hilbert action. Alternatively, we can also calculate the off-shell corrections by adding a boundary near the horizon to exclude the singularities on the horizon. Finally, the generalized free energy can be obtained from the reduced action using the Hawking-Gibbons gravitational path integral. The result is consistent with the thermodynamic definition and also provides a solid foundation for the recent studies on dynamical state transition~\cite{AR,AS,AT,AU,AV,AW,AX,AY,AZ,BA,BB,BC,BD,BE} and black hole topology~\cite{BG,BH,BI,BJ,BK,BL,BM,BN,BO,BP,BQ,BR,BS,BT,BU,BV,BW,BX,BY}.

Based on the generalized free energy landscape, we consider the dynamical state transition of the dyonic AdS black hole in the grand canonical ensemble as a stochastic process quantified by the Langevin equation. Unlike previous studies in the canonical ensemble with only one order parameter, we treat the horizon radius and the electric charge as the order parameters in the grand canonical ensemble. As the order parameter increases, the framework in the grand canonical ensemble differs from that in the canonical ensemble because of the difficulty in solving differential equations with boundary-value conditions. Consequently, we switch the dynamics from the time-dependent Newtonian description to the energy-dependent Hamilton-Jacobi description, allowing us to obtain the dominant path $Q(r)$ in the order parameter space with minimal action. We consider the state transition occurring along the dominant path $Q(r)$, and the two-dimensional dynamics can be transformed into one-dimensional dynamics. Finally, we use the path integral framework to calculate the time-dependent dominant path $r(t)$ and the rate of the state transition. The former shows how the process proceeds during the state transition, while the latter quantifies the time scale of the state transition. Moreover, we also derive the analytical expressions for the time evolution of the transition probabilities, which are found to be dependent on the transition rates between the small and large black hole states. These results are consistent with the qualitative analysis of the free energy landscape. Furthermore, because the time-independent dominant path in the order parameter space provided by the Hamilton-Jacobi description can always reduce the free order parameters to one dimension, our framework is applicable to ensembles with order parameters of arbitrary dimensions.

Finally, we discuss the applicability of our framework in the case of $q_M=0$. In this scenario, there are two black hole branches: the unstable small black hole and the stable large black hole. Moreover, the Hawking-Page phase transition can occur between the thermal AdS space and the stable large black hole. In our framework, we quantify the dynamics of state transition as a stochastic process described by the Langevin equation, with the driving force provided by the gradient of the generalized free energy. However, the generalized free energy is only defined for $r_h\geq 0$, and $r_h=0$ corresponds to the thermal AdS space state. Consequently, the driving force on the thermal AdS space state is not well-defined, and our framework cannot be applied when $q_M=0$. To overcome this limitation and ensure the applicability of our framework for $q_M=0$, further considerations are necessary to redefine the driving force on the thermal AdS space state.



\section*{Acknowledgments}
C. H. Liu would like to thank Hong Wang, He Wang, Xiaokun Yan, Linqi Wang, and Ligang Zhu for their helpful discussions. C. H. Liu also thanks the support from the National Natural Science Foundation of China Grant No. 21721003 and No. 12234019.

\appendix
\section{Counterterm subtraction} 
\label{Cts}
In the method of “counterterm subtraction”, the counterterm $I_{count}$ is added in the Euclidean action to cancel the divergence at infinity~\cite{CB,CC,CD,CE,CF}. Then, the Euclidean action is written as
\be
\ba
\label{eq:500}
I_{M/\Sigma_-}=I_{bulk}+I_{surf}+I_{count},
\ea
\ee
where 
\be
\ba
\label{eq:501}
I_{bulk}=-\frac{1}{16\pi}\int_{M/\Sigma_-}d^4x\sqrt{g}(R-2\Lambda-F^2),
\ea
\ee
\be
\ba
\label{eq:502}
I_{surf}=-\frac{1}{8\pi}\int_{\Sigma_+}d^3x\sqrt{h}K,
\ea
\ee
and 
\be
\ba
\label{eq:503}
I_{count}=\frac{1}{8\pi}\int_{\Sigma_+}d^3x\sqrt{h}[\frac{2}{l}+\frac{l}{2}R^{(3)}-\frac{l^3}{2}(R_{ab}^{(3)}R^{(3)ab}-\frac{3}{8}R^{(3)2})].
\ea
\ee
$h$ is the determinant of the induced metric $h_{\mu\nu}$ on the boundary $\Sigma_+$ at infinity, $K=h^{\mu\nu}K_{\mu\nu}$ is the trace of the extrinsic curvature of $\Sigma_+$ as embedded in $M$, $R^{(3)}$ and $R_{ab}^{(3)}$ are the Ricci scalar curvature and Ricci tensor for the boundary metric $h_{\mu\nu}$. We will calculate the Euclidean action for arbitrary temperature $T$, or equivalently, the period $\beta$ of imaginary time $\tau$ is not necessary to equal to $\beta_H=\frac{1}{T_H}$.

The calculation of $I_{bulk}$ is shown as
\be
\ba
\label{eq:504}
I_{bulk}&=\lim_{\epsilon\to 0}\frac{1}{16\pi}\int_0^{\beta}d\tau\int_{r_h+\epsilon}^{\Tilde{R}}dr\int_0^{\pi}d\theta\int_0^{2\pi}d\phi [r^2\sin\theta(\frac{6}{l^2}+2\frac{q_M^2-q_E^2}{r^4})]\\
&=\frac{\beta}{2l^2}(\Tilde{R}^3-r_h^3)+\frac{\beta(q_M^2-q_E^2)}{2r_h},
\ea
\ee
where we assume that the boundary $\Sigma_+$ locates at $r=\Tilde{R}$ and cut off the integral by $\Tilde{R}$. Finally, we will take the limit $\Tilde{R}\to\infty $.

Then, we will calculate $I_{surf}$. The nonvanishing components of the induced metric $h_{\mu\nu}$ are given by
\be
\ba
\label{eq:505}
h_{\tau\tau}=1+\frac{\Tilde{R}^2}{l^2}-\frac{2M}{\Tilde{R}}+\frac{q_E^2+q_M^2}{\Tilde{R}^2},
h_{\theta\theta}=\Tilde{R}^2,h_{\phi\phi}=\Tilde{R}^2\sin^2\theta.
\ea
\ee

In order to calculate $K$, we introduce the outpointing unit normal vector as $n^{\mu}=(0,\sqrt{1+\frac{r^2}{l^2}-\frac{2M}{r}+\frac{q_E^2+q_M^2}{r^2}},0,0)$. $K$ can be calculated as 
\be
\ba
\label{eq:506}
K&=h^{\mu\nu}K_{\mu\nu}=h^{\mu\nu}\nabla_{\mu}n_{\nu}=h^{\mu\nu}(\partial_{\mu}n_{\nu}-\Gamma^{\rho}_{\mu\nu}n_{\rho})\\
&=-(h^{\tau\tau}\Gamma^{r}_{\tau\tau}+h^{\theta\theta}\Gamma^r_{\theta\theta}+h^{\phi\phi}\Gamma^r_{\phi\phi})n_r|_{r=\Tilde{R}}\\
&=(\frac{2}{\Tilde{R}}+\frac{3\Tilde{R}}{l^2}-\frac{3M}{\Tilde{R}^2}+\frac{q_E^2+q_M^2}{\Tilde{R}^3})\frac{1}{\sqrt{1+\frac{\Tilde{R}^2}{l^2}-\frac{2M}{\Tilde{R}}+\frac{q_E^2+q_M^2}{\Tilde{R}^2}}}.
\ea
\ee

By substituting Eq.~(\ref{eq:506}) into Eq.~(\ref{eq:502}), we can obtain
\be
\ba
\label{eq:507}
I_{surf}=-\frac{\beta}{2}(2\Tilde{R}+\frac{3\Tilde{R}^3}{l^2}-3M+\frac{q_E^2+q_M^2}{\Tilde{R}}).
\ea
\ee

Then, we calculate $I_{count}$ as
\be
\ba
\label{eq:508}
I_{count}&=\frac{1}{8\pi}\int_0^{\beta}d\tau\int_0^{\pi}d\theta\int_0^{2\pi}d\phi
\sqrt{1+\frac{\Tilde{R}^2}{l^2}-\frac{2M}{\Tilde{R}}+\frac{q_E^2+q_M^2}{\Tilde{R}^2}}\Tilde{R}^2\sin\theta(\frac{2}{l}+\frac{l}{\Tilde{R}^2}-\frac{l^3}{4\Tilde{R}^4})\\
&=\frac{\beta\Tilde{R}^3}{l^2}+\beta\Tilde{R}-\beta M,
\ea
\ee
where we have used $R^{(3)}=\frac{2}{\Tilde{R}^2}$, $R_{ab}^{(3)}R^{(3)ab}=\frac{2}{\Tilde{R}^4}$ and finally abandoned all the inverse terms of $\Tilde{R}$ due to $\Tilde{R}\to\infty$. During the calculations, the Taylor expansion has been used as 
\be
\ba
\label{eq:509}
\sqrt{1+\frac{\Tilde{R}^2}{l^2}-\frac{2M}{\Tilde{R}}+\frac{q_E^2+q_M^2}{\Tilde{R}^2}}
=\frac{\Tilde{R}}{l}[1+\frac{l^2}{2\Tilde{R}^2}-\frac{l^2M}{\Tilde{R}^3}+O(\frac{1}{\Tilde{R}^4})].
\ea
\ee

Finally, the Euclidean action can be calculated as
\be
\ba
\label{eq:510}
I_{M/\Sigma_-}&=I_{bulk}+I_{surf}+I_{count}\\
&=\frac{\beta M}{2}-\frac{\beta r_h^3}{2l^2}+\frac{\beta(q_M^2-q_E^2)}{2r_h}\\
&=\frac{\beta r_h}{4}-\frac{\beta r_h^3}{4l^2}-\frac{\beta q_E^2}{4r_h}+\frac{3\beta q_M^2}{4r_h}.
\ea
\ee


\section{Background subtraction}
\label{Bs}
The metric $g'_{\mu\nu}$ of the AdS space takes the same form as Eq.~(\ref{eq:3}), however, $f(r)$ now becomes 
\be
\ba
\label{eq:511}
f(r)=1+\frac{r^2}{l^2}.
\ea
\ee

We should match the metric of the AdS space to that of the dyonic AdS black hole at a cutoff distance $\Tilde{R}$, and we have
\be
\ba
\label{eq:512}
(1+\frac{\Tilde{R}^2}{l^2}-\frac{2M}{\Tilde{R}}+\frac{q_E^2+q_M^2}{\Tilde{R}^2})d\tau^2=(1+\frac{\Tilde{R}^2}{l^2})d\tau'^2.
\ea
\ee
This in turn gives the relationship between the time period $\beta'$ of the AdS space and the time period $\beta$ of the dyonic AdS black hole as
\be
\ba
\label{eq:513}
\beta'&=\beta(\frac{1+\frac{\Tilde{R}^2}{l^2}-\frac{2M}{\Tilde{R}}+\frac{q_E^2+q_M^2}{\Tilde{R}^2}}{1+\frac{\Tilde{R}^2}{l^2}})^{\frac{1}{2}}\\
&=\beta[1-\frac{Ml^2}{\Tilde{R}^3}+O(\frac{1}{\Tilde{R}^4})].
\ea
\ee

Finally, the Euclidean action can be calculated as 
\be
\ba
\label{eq:514}
I_{M/\Sigma_-}&=-\frac{1}{16\pi}\int_{M/\Sigma_-}d^4x\sqrt{g}(R-2\Lambda-F^2)+\frac{1}{16\pi}\int d^4x\sqrt{g'}(R-2\Lambda)\\
&=\frac{\beta}{4}\lim_{\epsilon\to 0}\int_{r_h+\epsilon}^{\Tilde{R}}dr[\frac{6r^2}{l^2}+\frac{2(q_M^2-q_E^2)}{r^2}]-\frac{3\beta'}{2l^2}\int_0^{\Tilde{R}}drr^2\\
&=\frac{\beta M}{2}-\frac{\beta r_h^3}{2l^2}+\frac{\beta(q_M^2-q_E^2)}{2r_h}\\
&=\frac{\beta r_h}{4}-\frac{\beta r_h^3}{4l^2}-\frac{\beta q_E^2}{4r_h}+\frac{3\beta q_M^2}{4r_h},
\ea
\ee
where Eq.~(\ref{eq:513}) is used to replace $\beta'$ with $\beta$, and we should note that the lower limit of $r$ in the integral of the even-dimensional AdS space is always $0$~\cite{CI}. In a word, we have obtained the same result with the method of “counterterm subtraction”.

\section{A possible approach for calculating the off-shell corrections}
\label{Apa}
In the main body of the paper, we have calculated the off-shell corrections by regularizing the divergences of the Ricci scalar curvature $R$ and the electromagnetic potential $A_{\mu}$ at the horizon. Alternatively, we can also exclude some regions of spacetime near the horizon by adding a boundary near the horizon. Recalling the Euclidean path integral, Eq.~(\ref{eq:17}) indicates that $\rho=0$ corresponds to $r=r_h$, which means that $r$ takes values greater than or equal to the radius of the horizon. In the regular manifold with no singularity, the Euclidean section is bounded by the surface $\Sigma_+$ at infinity. However, our Euclidean section is singular at the horizon, and the boundary should be taken as $\Sigma_-$ near the horizon and $\Sigma_+$ at infinity.

At first, we calculate the off-shell correction resulting from $T\neq T_H$. After we add a boundary $\Sigma_-$ near the horizon, in order to have a well-defined action principle, we should add a new Gibbons-Hawking boundary term:
\be
\ba
\label{eq:1111}
I_{\Sigma_-}^c&=-\frac{1}{8\pi}\int_{\Sigma_-}d^3x\sqrt{h}K\\
&=\lim_{\epsilon\to 0}\frac{1}{8\pi}\int_0^{\beta}d\tau\int_{0}^{\pi}d\theta\int_0^{2\pi}d\phi\{{(r_h+\epsilon)}^2\sin{\theta}[\frac{2}{r_h+\epsilon}+\frac{3{(r_h+\epsilon)}}{l^2}-\frac{3M}{{(r_h+\epsilon)}^2}+\frac{q_E^2+q_M^2}{{(r_h+\epsilon)}^3}]\}\\
&=\frac{\beta}{\beta_H}\pi r_h^2,
\ea
\ee
where $K=h^{\mu\nu}K_{\mu\nu}=h^{\mu\nu}\nabla_{\mu}n_{\nu}$ and $n_{\mu}$ is the inward pointing unit normal vector for $\Sigma_-$. If $\beta=\beta_H$, the reduced action should recover the on-shell action $I_{os}$. Therefore, another term $I_{\Sigma_1}^c$ needs to be added such that $I_{\Sigma_-}^c+I_{\Sigma_1}^c=0$ when $\beta=\beta_H$. $I_{\Sigma_1}^c$ can be taken as
\be
\ba
\label{eq:1112}
I_{\Sigma_1}^c&=-\frac{1}{4}\int_0^{\pi}d\theta\int_0^{2\pi}d\phi\sqrt{h_{1}}\\
&=-\pi r_h^2,
\ea
\ee
where $\Sigma_1$ is a two-dimensional sphere near the horizon and $h_1$ is the induced metric on $\Sigma_1$. When we consider the variation of $I_{\Sigma_1}^c$, we can find that $I_{\Sigma_1}^c$ does not contribute to the equations of motion for the gravitational field and electromagnetic field. Namely, this term will not affect the well-defined action principle. If we sum $I_{\Sigma_-}^c$ and $I_{\Sigma_1}^c$ for arbitrary $\beta$, the off-shell correction $I_{cc}$ in Eq.~(\ref{eq:27}) will be recovered.

Then, we calculate the off-shell correction resulting from $\Phi_E\neq\Phi_{EH}$. Specifically, we add the boundary $\Sigma_-$ to exclude the singularity of $A_{\mu}$ and impose the fixed electric charge condition on $\Sigma_-$. It means another term is added to the Euclidean action as~\cite{CP} 
\be
\ba
\label{eq:37}
I_{ce}=-\frac{1}{4\pi}\int_{\Sigma_-}d^3x \sqrt{h_-}{F^{\mu\nu}}n_{\mu}A_{\nu}.
\ea
\ee
Then, the boundary-related terms induced by the variation of the total action with respect to $A_{\mu}$ are
\be
\ba
\label{eq:38}
\frac{1}{4\pi}\int_{\Sigma_+}d^3x \sqrt{h_+}n_{\mu}F^{\mu\nu}\delta A_{\nu}-\frac{1}{4\pi}\int_{\Sigma_-}d^3x A_{\nu}\delta(\sqrt{h_-}F^{\mu\nu}n_{\mu}),
\ea
\ee
where $h_+$ and $h_-$ are the induced metrics on $\Sigma_+$ and $\Sigma_-$, and $n_{\mu}$ is the outward pointing unit normal vector for $\Sigma_+$ and inward pointing unit normal vector for $\Sigma_-$. If we use the Maxwell equation, we can find that $\delta(\sqrt{h_-}F^{\mu\nu}n_{\mu})$ is equivalent to $\delta q_E$. Thus, Eq.~(\ref{eq:38}) implies that the electric potential $\Phi_E$ is fixed on $\Sigma_+$, and the electric charge $q_E$ is fixed on $\Sigma_-$. The calculations of $I_{ce}$ in Eq.~(\ref{eq:37}) yields
\be
\ba
\label{eq:39}
I_{ce}=\beta q_E(\frac{q_E}{r_h}-\Phi_E),
\ea
\ee
so the same result as Eq.~(\ref{eq:35}) has been obtained.

In conclusion, we can obtain the same generalized free energy as Eq.~(\ref{eq:522}) in the main body of the paper by adding a boundary near the horizon.

\section{A simple example for showing the different characteristic time scales}
\label{Asef}
When the friction coefficient $\gamma$ is very large, the inertia term can be neglected, regardless of the specific form of the driving force provided by the thermodynamic potential\cite{QAA, QAB, QAC, CR}. Without loss of generality, we use the one-dimensional harmonic oscillator potential as a simple example to analyze the characteristic time scales of the variables. The deterministic part of the Langevin equation can be rewritten as:
\be
\ba
\label{eq:7123}
\frac{dv}{dt}=-\gamma v-x,
\ea
\ee
\be
\ba
\label{eq:7124}
\frac{dx}{dt}=v.
\ea
\ee

We will show that a large $\gamma$ value leads to significantly different time scales for the relaxation of $v$ in Eq.~(\ref{eq:7123}) and $x$ in Eq.~(\ref{eq:7124}). At first, we can solve Eq.~(\ref{eq:7123}) and Eq.~(\ref{eq:7124}) for $\gamma>2$, which yields
\be
\ba
\label{eq:7215}
x(t)=A_1\exp{[\lambda_1 t]}+A_2\exp{[\lambda_2 t]},
\ea
\ee
\be
\ba
\label{eq:7216}
v(t)=A_1\lambda_1\exp{[\lambda_1 t]}+A_2\lambda_2\exp{[\lambda_2 t]},
\ea
\ee
where
\be
\ba
\label{eq:7217}
\lambda_1=\frac{-\gamma+\sqrt{\gamma^2-4}}{2}, \qquad \lambda_2=\frac{-\gamma-\sqrt{\gamma^2-4}}{2},
\ea
\ee
and $A_1$, $A_2$ are determined by the initial conditions.

\begin{figure}[t]
\centering
\includegraphics[width=0.68\textwidth]{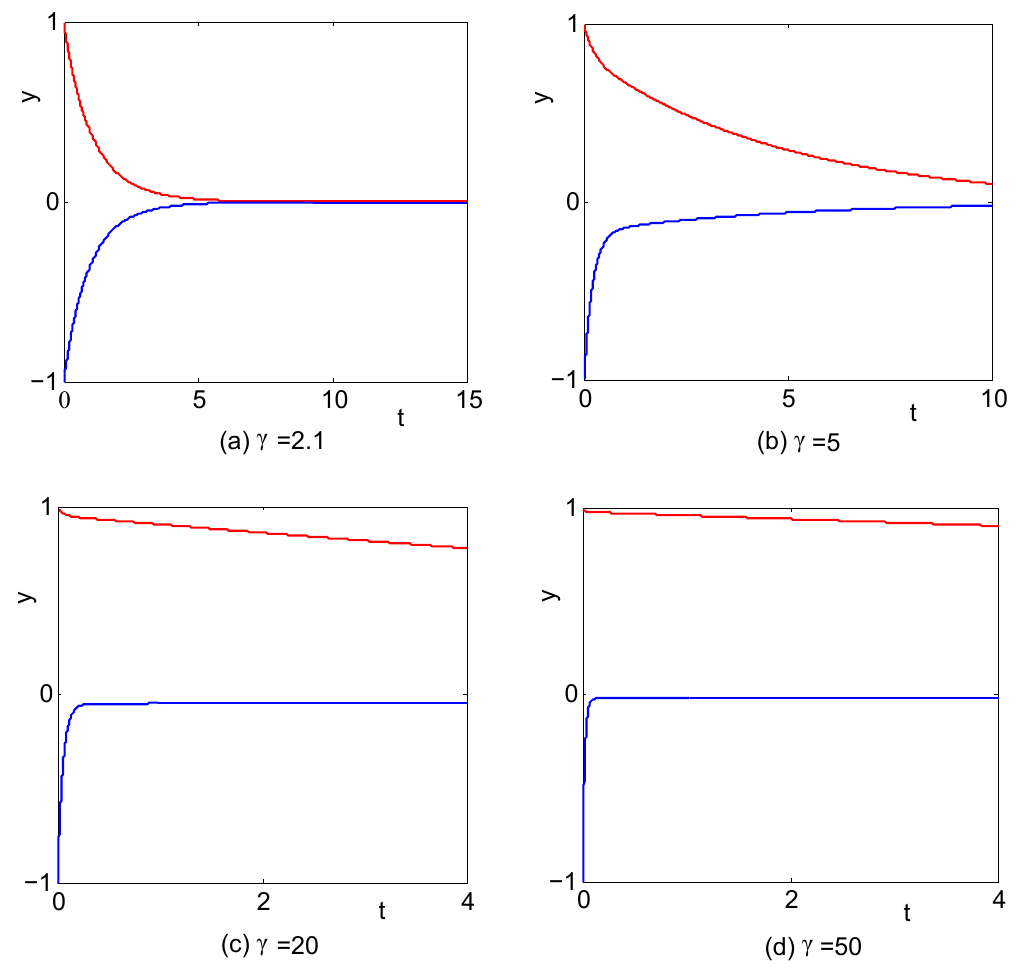}
\caption{The figures show $x(t)$ and $v(t)$ as the solutions of Eq.~(\ref{eq:7123}) and Eq.~(\ref{eq:7124}) for various $\gamma$ values. The red lines indicate $x(t)$, while the blue lines represent $v(t)$. The vertical coordinate $y$ represents the values of $x(t)$ and $v(t)$, and the initial conditions are $x(0)=1$ and $v(0)=-1$. Eventually, both $x(t)$ and $v(t)$ converge to a steady value of $0$.}
\label{Fig120}
\end{figure}

In Fig.~\ref{Fig120}, we have plotted $x(t)$ and $v(t)$ for different $\gamma$ values. Both $x(t)$ and $v(t)$ eventually converge to a steady value of $0$. As $\gamma$ increases, the time for $v$ to reach the steady value is shorter compared to $x$. When $\gamma$ is very large, the characteristic time scales of $v$ and $x$ differ significantly, with $v$ being the fast variable and $x$ being the slow variable.

\end{document}